\documentclass[12pt,preprint]{aastex}

\newcommand\pac{Paczy{\'n}ski }

\newcommand\msun {M_\odot}

\newcommand\mearth {{M_\oplus}}

\usepackage{lscape}
\usepackage{comment}

\shorttitle{Cold Planet Mass Ratio Function from MOA-II}
\shortauthors{Suzuki et al.}

\newcommand\ltsima{$\; \buildrel <\over\sim \;$}
\newcommand\simlt{\lower.5ex\hbox{\ltsima}}
\newcommand\gtsima{$\; \buildrel >\over\sim \;$}
\newcommand\simgt{\lower.5ex\hbox{\gtsima}}

\begin{document}


\title{The Exoplanet Mass-Ratio Function from the MOA-II Survey:\\
Discovery of a Break and Likely Peak at a Neptune Mass}

\author{D.~Suzuki\altaffilmark{1}, 
D.P.~Bennett\altaffilmark{1,2}, 
T.~Sumi\altaffilmark{3},
 I.A.~Bond\altaffilmark{4},
L.A.~Rogers\altaffilmark{5,14},
F.~Abe\altaffilmark{6}, 
Y.~Asakura\altaffilmark{6}, 
A.~Bhattacharya\altaffilmark{1,2}, 
M.~Donachie\altaffilmark{7}, 
M.~Freeman\altaffilmark{7}, 
A.~Fukui\altaffilmark{8}, 
Y.~Hirao\altaffilmark{3}, 
Y.~Itow\altaffilmark{6},
N.~Koshimoto\altaffilmark{3}, 
M.C.A.~Li\altaffilmark{7},
C.H.~Ling\altaffilmark{4}, 
K.~Masuda\altaffilmark{6}, 
Y.~Matsubara\altaffilmark{6}, 
Y.~Muraki\altaffilmark{6}, 
M.~Nagakane\altaffilmark{3}, 
K.~Onishi\altaffilmark{9}, 
H.~Oyokawa\altaffilmark{6}, 
N.~Rattenbury\altaffilmark{7}, 
To.~Saito\altaffilmark{10}, 
A.~Sharan\altaffilmark{8}, 
H.~Shibai\altaffilmark{3}, 
D.J.~Sullivan\altaffilmark{11}, 
P.J.~Tristram\altaffilmark{12}, and 
A.~Yonehara\altaffilmark{13} \\
(the MOA collaboration)\\
}


\altaffiltext{1}{Laboratory for Exoplanets and Stellar Astrophysics, NASA/Goddard Space Flight Center, Greenbelt, MD 20771, USA}
\altaffiltext{2}{Department of Physics, University of Notre Dame, Notre Dame, IN 46556, USA}
\altaffiltext{3}{Department of Earth and Space Science, Graduate School of Science, Osaka University, 1-1 Machikaneyama, Toyonaka, Osaka 560-0043, Japan}
\altaffiltext{4}{Institute of Information and Mathematical Sciences, Massey University, Private Bag 102-904, North Shore Mail Centre, Auckland, New Zealand}
\altaffiltext{5}{Department of Astronomy \& Astrophysics, University of Chicago, 5640 S Ellis Ave, Chicago, IL 60637, USA}
\altaffiltext{6}{Institute for Space-Earth Environmental Research, Nagoya University, Furo-cho, Chikusa, Nagoya, Aichi 464-8601, Japan}
\altaffiltext{7}{Department of Physics, University of Auckland, Private Bag 92019, Auckland, New Zealand}
\altaffiltext{8}{Okayama Astrophysical Observatory, National Astronomical Observatory, 3037-5 Honjo, Kamogata, Asakuchi, Okayama 719-0232, Japan}
\altaffiltext{9}{Nagano National College of Technology, Nagano 381-8550, Japan}
\altaffiltext{10}{Tokyo Metropolitan College of Industrial Technology, Tokyo 116-8523, Japan}
\altaffiltext{11}{School of Chemical and Physical Sciences, Victoria University, Wellington, New Zealand}
\altaffiltext{12}{Mt. John University Observatory, P.O. Box 56, Lake Tekapo 8770, New Zealand}
\altaffiltext{13}{Department of Physics, Faculty of Science, Kyoto Sangyo University, Kyoto 603-8555, Japan}
\altaffiltext{14}{Sagan Fellow, Department of Earth and Planetary Science, University of California at Berkeley, 501 Campbell Hall \#3411, Berkeley, CA 94720, USA}


\begin{abstract}
We report the results of the statistical analysis of planetary signals discovered in 
MOA-II microlensing survey alert system events from 2007 to 2012.
We determine the survey sensitivity as a function of planet-star mass ratio, $q$, 
and projected planet-star separation, $s$, in Einstein radius units.
We find that the mass ratio function
is not a single power-law, but has a change in slope at $q \sim 10^{-4}$, 
corresponding to $\sim 20\mearth$ for the median host star mass of $\sim 0.6 \msun$.
We find significant planetary signals in 23 of the 1474 alert events
that are well characterized by the MOA-II survey data alone.
Data from other groups are used only to characterize planetary signals that have been 
identified in the MOA data alone.
The distribution of mass ratios and separations
of the planets found in our sample are well fit by a broken power-law model of the form
$dN_{\rm pl}/(d\log q\ d\log s) = A (q/q_{\rm br})^n s^m \, {\rm dex}^{-2}$ for $q >  q_{\rm br}$ and 
$dN_{\rm pl}/(d\log q\ d\log s) = A (q/q_{\rm br})^p s^m \, {\rm dex}^{-2}$ for $q < q_{\rm br}$,
where $q_{\rm br}$ is the mass ratio of the break. We also combine this analysis with the
previous analyses of Gould et al. and Cassan et al., bringing the total sample to 30 planets.
This combined analysis yields
$A = 0.61^{+0.21}_{-0.16}$, $n =-0.93\pm 0.13$, $m = 0.49_{-0.49}^{+0.47}$ and 
$p = 0.6^{+0.5}_{-0.4}$ for $q_{\rm br}\equiv 1.7\times 10^{-4}$. 
The unbroken power law model is disfavored with a $p$-value
of 0.0022, which corresponds to a Bayes factor of 27 favoring the broken power-law
model.
These results imply that cold Neptunes are likely to be the
most common type of planets beyond the snow line.
\end{abstract}





\section{Introduction}
\label{sec_intro}

More than 50 exoplanets have been discovered by the gravitational microlensing
method \citep{mao91}, since the first discovery more than a decade ago \citep{bon04}. 
This is a small fraction of the number of
planets found by the transit \citep{kepler_q16} and radial velocity (RV; Butler et al. 2006) methods,
but the microlensing method can detect a class of planets that are invisible to other
methods. Microlensing is uniquely sensitive to planets down to an Earth-mass 
\citep{ben96} beyond the snow line, where the other detection methods are ineffective 
\citep{bennett_rev,gau12}. According to the core accretion theory \citep{lissauer_araa,ida04},
the snow line is the radius in the protoplanetary disk beyond which the temperature 
is cold enough for ices to condense. 
The presence of ices increases the density of
solids in the protoplanetary disk by a factor of 
3--4 \citep{kk08}, and this is thought to allow the ice-rock cores of giant planets to form more
quickly than solid planets can condense in the inner parts of protoplanetary disks. 
The core accretion theory predicts \citep{ida04} 
that gas giant planet cores can grow to $\sim10$ Earth-masses of solid material beyond
the snow line, and this is enough to trigger the rapid accretion of hydrogen and helium
gas needed to form a gas giant. However, there is a danger that the hydrogen and helium
might dissipate before the core grows larger enough to trigger rapid accretion. In this case,
we are likely to be left with a failed gas-giant-core planet. It is thought that these 
failed gas-giant core planets may be especially common around low-mass host stars
\citep{laughlin04}. The early results from microlensing support this idea. Microlensing
has found numerous examples of planets that are consistent with the expectations
for these failed gas-giant-core planets \citep{jp06,gou06,mur11,fur13,gou14,ben15hst,bat15}.
Other methods have not been able to detect such planets, so
the microlensing method is uniquely sensitive to this population of failed gas-giant-core planets.

The microlensing method is also uniquely sensitive to old, free-floating planets that
may have been ejected from the planetary system of their formation. While it is possible
to detect young, unbound planetary mass objects directly in the infrared \citep{delhome12,beichman13},
the infrared surveys do not yet have the sensitivity to detect planets of $\sim 1$ Jupiter-mass,
where the signal of a large free-floating, or rogue, planet population has been found by
microlensing \citep{sum11}. Some have suggested that this population of apparently rogue
planets might be bound in orbits so wide that microlensing cannot detect the host stars
\citep{quanz12}, but the microlensing results indicate that the median projected distance from
rogue planets to potential host stars is likely to be $>30\,$AU \citep{bennett12}.

There have been several previous statistical studies of exoplanets found by microlensing.
Statistical studies \citep{gau02,sno04} from early microlensing survey data with no planet detections
were able to put upper limits on only the cold gas giant planet frequency.
Using 10 microlensing planet detections, \citet{sum10} looked at the distribution of 
planets as a function of their mass ratio, $q$,
and found that the exoplanet distribution beyond the snow line could be explained by
a power law mass ratio function of the form, $dN/d\,\log q = q^{-0.68 \pm 0.20}$, 
but this calculation was done without a calculation of the absolute exoplanet detection efficiency. 
(Throughout this paper, we use $\log$ to refer to the base-10 logarithm.)
They used the detection efficiency for each planetary event, but did not calculate the
efficiency for events without planetary signals. This allowed the determination of the
relative efficiency for different types of planets, but it was insufficient to determine the 
absolute detection efficiency. So, \citet{sum10} could only calculate the power-law index of 
the mass ratio function, but not its normalization.
\citet{gou10} took advantage of the very high detection efficiency provided by high
magnification microlensing events \citep{griest98,rhi2000} to do the first calculation of the
absolute detection efficiency with a sample of 13 high magnification microlensing events including 6 planets.
While the very high planet detection efficiency is provided by high magnification events, 
there is a well-known degeneracy between close and wide planet solutions \citep{dom99,chu05} that
implies that, for most high magnification events, there is a significant ambiguity in the
star-planet separation unless it is close to the Einstein radius. As a result, it is difficult to measure
the separation dependence of the planet population with a sample of high magnification
events, and \cite{gou10} did not attempt to do so. They found a planet frequency of
$d^{2}N/d\,\log q\,\,d\,\log s = 0.36 \pm 0.15$ at $q = 5\times 10^{-4}$ beyond 
the snow line.

The PLANET collaboration \citep{cas12} also published a statistical analysis of planetary
events found in their 2002--2007 data. They had a much larger sample of microlensing
events, but the number of planets discovered was smaller, 3, compared to 6 planets found
in the \citet{gou10} sample. Because one planet is common to both samples, the
combined sample includes 8 planets, and \citet{cas12} use this combined sample and the 
slope of the power law mass ratio function from \citet{sum10} to find
the cold planet mass function 
$d^{2}N/d\,\log M\,\,d\,\log a =  0.24^{+0.16}_{-0.10}(M/M_{\rm Sat})^{-0.73\pm0.17}$, 
where $M_{\rm Sat} = 95.2 M_{\oplus}$. This result is quoted in terms of planetary mass 
instead of mass ratio because \citet{cas12} used a Bayesian method 
\citep{dom06} to convert from the
mass ratios, $q$, measured in the microlensing light curves to masses. This method 
implicitly assumes that the probability for a star to host a planet of a given mass ratio 
is independent of the stellar mass and location of the system through the Galaxy. 
This assumption is probably safe with such a small
sample, but it could be problematic with samples large enough to measure 
the dependence of the exoplanet mass function on the mass of and distance to the host star.

A more recent paper is based on three-site survey data using combined data from the
Optical Gravitational Lensing Experiment (OGLE), MOA, and Wise wide-field microlensing surveys \citep{shv15}. This has some overlap with
our own sample, as it includes data from 2011 to 2014, but it only includes events from
the $8\,{\rm deg}^2$ covered by all three surveys. 
They used a coarse grid in mass ratio and separation for the light curve modeling of 
some of the binary and planetary events found in their sample. This appears to have
resulted in important errors in the parameters for some events, as we discuss in 
Section~\ref{subsec:com_micro}.
Nevertheless, they estimate a slope of  $dN/d\,\log q = q^{-0.50 \pm 0.17}$.
They also estimated the total exoplanet frequency for $10^{-4.9} < q < 10^{-1.4}$. 

Recent studies \citep{cla14a,cla14b,mon14} have compared the exoplanet distribution found 
by microlensing with the results of RV observations of M-dwarfs and
found that the results from both methods are consistent, although the radial 
velocity is only sensitive to planets of Jupiter-mass or greater beyond the snow line.

The {\it Kepler} telescope \citep{bor10,kepler_q16} has revealed the planet population 
in close orbits, mostly within 1 AU around solar type stars.
Most of detected planets or candidates are located within the orbit of Mercury and many 
multiple planets systems consisting of super Earths have been found \citep{how12}.
Microlensing also has detected a number of super Earths \citep{jp06,gou06,ben08,mur11,gou14}, 
but the planets found by microlensing are in much longer period orbits around somewhat lower-mass stars
than the small planets found by {\it Kepler}. Also, the masses for small {\it Kepler} planets 
with longer orbital periods are rarely determined,
and when they are determined, they show
surprisingly low densities \citep{lis13}. Attempts to measure a mass-radius relation find a scatter
much larger than the measurement errors \citep{wei14,wol15}, indicating a range of compositions
for exoplanets at a fixed mass or radius.
The average exoplanet composition could change dramatically between the well measured 
{\it Kepler} planets at $\ll 1\,$AU and microlens planets beyond the snow line. 
Thus, it is difficult to compare statistical analyses of exoplanet frequency from {\it Kepler} data with 
those from current, ground-based microlensing survey data.
Future space-based exoplanet microlensing surveys
\citep{ben02,pen13}, such as the {\it WFIRST} survey \citep{spe15}, 
will explore a much wider range of orbital periods and will therefore have overlap with
{\it Kepler}.

The one observable that relates to interesting lens properties in single mass microlensing
light curves is the event time scale, the time required for a source star to cross the angular Einstein radius,
\begin{equation}
t_{\rm E} = \frac{\theta_{\rm E}}{\mu_{\rm rel}}, ~~~~~ \theta_{\rm E } = \sqrt{{4GM\over c^2} \left(\frac{1}{D_{\rm L}} - \frac{1}{D_{\rm S}}\right)},
\end{equation}
where $\mu_{\rm rel}$ is the relative lens-source proper motion, $M$ is mass of the lens system, 
and $D_{\rm L}$ and $D_{\rm S}$ are the distance to the lens and source star, respectively.
In planetary microlensing events, where signals of both the planet and its host star are seen,
we can also measure planet-star mass ratio and angular separation relative to the angular Einstein radius.
In most planetary events, the sharp planetary light curve features resolve the finite angular
size of the source star, allowing the source radius crossing time, $t_*$, to be measured. 
Since the angular radius of the source star, $\theta_*$, can be determined from its 
brightness and color \citep{boy14}, the measurement of $\theta_*$ also provides a
measurement of the angular Einstein radius, $\theta_{\rm E} = \theta_* t_{\rm E}/t_*$.
Unfortunately, this is not enough to determine the mass or distance of the lens system.
Instead, we are left with the following mass-distance relation,
\begin{equation}
M = {c^2\over 4G} \theta_E^2 {D_S D_L\over D_S - D_L} 
       = 0.9823\,\msun \left({\theta_E\over 1\,{\rm mas}}\right)^2\left({x\over 1-x}\right)
       \left({D_S\over 8\,{\rm kpc}}\right) \ ,
\label{eq-m_thetaE}
\end{equation}
where $x = D_L/D_S$. In some cases, the actual mass can be determined if a 
microlensing parallax signal \citep{gould-par1,alc95} is measured 
\citep{ben08,gau08,mur11,fur13,ben16cir} or when the
host star is detected with high angular resolution adaptive optics 
\citep{bat14,bat15} or Hubble Space Telescope (HST) \citep{ben06,ben07,ben15hst} observations, but in
most cases, the mass and distance to the lens are not directly determined. Often, microlens planet
discovery papers estimate the mass and distance with a Bayesian analysis, under the assumption
that exoplanet properties do not depend on the host mass and distance. We would eventually
like to measure this dependence on the host mass and distance, so we cannot use such mass 
estimates that assume the answer to a question that we are trying to investigate.

In this paper, we report on our analysis of the exoplanet mass ratio function as a function of
separation (in the angular Einstein radius units) using microlensing events detected by the 
Microlensing Observations in Astrophysics (MOA) collaboration alert system
during the years 2007-2012. 
We use planet to host star mass ratio instead of planet mass, because the former is measurable 
without any assumption regarding the estimate of host star masses.
This is the first statistical analysis of the prevalence and properties of exoplanets based on 
microlensing survey data, which includes full light curve analysis of binary and 
planetary events to precisely determine the mass ratio and separation for events with
significant planetary or binary signals.
Previous analyses based on microlensing survey data either identified no definitive planetary events
\citep{sno04,tsa03,tsa16}, or included an incomplete light curve analysis that did not  
definitively identify planetary events \citep{shv15}. Previous complete statistical 
analyses \citep{gou10,cas12} have been based on planets detected in the 
data from microlensing follow-up groups who focus on events thought to have relatively high
sensitivity to exoplanets. In contrast, microlensing survey groups use large field-of-view telescopes
to observe a much larger number of events. Many of these events have low planet detection
efficiency, but overall, the sensitivity of the surveys is higher because of the much larger
number of events observed \citep{tsa16}. This is why the new microlensing projects just coming online
now \citep{kim16kmtnet} and planned for the future \citep{spe15} are all high cadence surveys
that will find exoplanet signals in high cadence, wide-field, survey data.

This paper is organized as follows.
We describe the MOA survey observations, as well as observations by other telescopes 
that are used to help classify and characterize anomalous microlensing events
in Section \ref{sec_obs}. In Section \ref{sec_eve_sel}, we explain how microlensing events
from the full 2007-2012 sample are selected for use to search for planetary signals and determine
our planet detection efficiency.
We describe our detection efficiency to planets in Section \ref{sec_est_de}.
The exoplanet mass ratio function is presented in Section \ref{sec_pla_fre}, and our 
results are discussed in Section \ref{sec_dis}.
Finally, we summarize our results in Section \ref{sec_sum}.

\section{Observations}
\label{sec_obs}

The microlensing method requires the monitoring of tens of millions of stars to find gravitational
microlensing events. It also requires high cadence photometric monitoring to detect 
planetary signals, which have short time scales, ranging from a few hours to a few days
for Earth and Jupiter-mass planets, respectively.
To find such rare, unpredictable planetary signals, the survey and follow-up teams 
have collaborated to comprise a large observational network.

Since 2006, the Microlensing Observations in Astrophysics (MOA)
collaboration has surveyed the Galactic bulge using the dedicated 
1.8\,m MOA-II telescope at Mt.John University Observatory in New Zealand. This telescope 
employs the MOA-Cam3 \citep{sak08} CCD camera consisting of ten 
$2\,{\rm k}\times 4\,{\rm k}$ detectors with a plate scale of 0.58 arcsec ${\rm pixel}^{-1}$. It is 
mounted at the prime focus and has a $2.2\,{\rm deg}^{2}$ field-of-view (FOV).
This wide FOV allows us to monitor $\sim\!44\,{\rm deg}^{2}$ of the Galactic bulge 
with cadences ranging from 15 minutes for the fields with the highest microlensing rate,
to 45 minutes for fields with a medium microlensing rate,
to 95 minutes for the fields with the lowest microlensing rate. These high cadence observations
allow the MOA survey to detect planetary signals in all the observed microlensing events, rather
than just the fraction of events that are monitored for a fraction of their duration 
in the survey plus follow-up strategy
suggested by \citet{gou92} and employed in the previous statistical analysis of
\citet{gou10}\footnote{They used 13 high magnification events from 4-years of observations.} 
and \citet{cas12}\footnote{They used 199 events from 6-years of observations.}. 
The larger number of events that can be surveyed for
planets leads to a higher planet detection rate. We find nearly 4 planetary events per
year compared to an average of about 1 planetary event per year from the alert and 
follow-up strategy.

Once MOA identifies a candidate microlensing event in progress, we issue a microlensing
event alert\citep{bon01}, often within 1 hour of the observation that triggered the alert.
It is these events, found by the MOA alert system that we use for our sample of microlensing
events. Our sample includes events announced by the MOA alert system in 2007 -- 2012, and
they consist of 488, 477, 563, 607, 485, and 680 events, respectively in the
years 2007 -- 2012. The MOA alert system was only partially functional in 2006, so the 2006
data are not included in this analysis. 

The other major survey group that operated during the 2007 -- 2012 seasons is the
Optical Gravitational Lensing Experiment
(OGLE) collaboration \citep{ogle4} which uses the 1.3 m Warsaw Telescope at the Las Campanas 
Observatory in Chile. The years 2007, 2008, and very early 2009 were the final years of the
OGLE-III survey using a CCD camera with a field-of-view of $0.25\,{\rm deg}^2$, which
was too small a field-of-view for a high cadence survey, but in 2010, OGLE upgraded its
CCD camera to the $1.4\,{\rm deg}^2$ OGLE-IV camera and began the high cadence 
OGLE-IV survey.

A third high cadence survey was started in 2011 at the Wise Observatory in Israel in order 
to fill the longitude gap between the MOA-II and OGLE-IV \citep{shv12}. The Wise survey uses a 
$1.0\,{\rm deg}^2$ CCD camera mounted on the Wise 1.0m telescope. Because of the
Wise Observatory's latitude of $30.60^\circ\,$N, the Wise survey can observe the bulge for
no more than 5 hours per night. As a result, they have limited their high cadence survey 
observations to $8\,{\rm deg}^2$ of the Galactic bulge.

In addition to these survey teams, most of the planetary and high magnification events are 
observed by the microlensing follow-up groups, such as the
Probing Lensing Anomalies NETwork (PLANET), the $\mu$lensing Follow-up 
Network ($\mu$FUN), RoboNet and the Microlensing Network for the Detection of 
Small Terrestrial Exoplanets (MiNDSTEp). Originally, before the high cadence surveys
came online, these follow-up programs were needed to get high cadence observations
of the events thought to be most promising for planet detection, following the alert plus
follow-up strategy laid out by \citet{gou92}. This strategy led to the discovery of number of  
high-magnification planetary events, \citep[e.g.]{gou06,don09mb07400,jan10,miy11,yee12,bac12pl}, 
as well as a couple of planetary signals in low-magnification events \citep{jp06,bat11}.

Follow-up observations were also used to help characterize planetary signals that were
detected in progress in the MOA-II survey data \citep{sum10,mur11,fur13,sko14}. It is 
important to treat such follow-up observations carefully in a statistical analysis. 
If these observations are triggered by the detection of a planetary signal, they cannot be used
to increase the significance of the planetary signal above the planet detection threshold, 
because they were only taken because the planetary signal was already detected.
The situation is different with the additional observations that are triggered by a high
magnification alert. Many high magnification events receive intensive observations
from follow-up groups because they have much more sensitivity to planetary signals than
low-magnification events. The crucial difference is that the additional observations for
a high magnification events are triggered by this {\it high sensitivity} and not an actual
planetary signal. So, very high cadence observations of high magnification events
are often planned before any planetary signal is detected, so it is possible to use
the follow-up observations to set detection thresholds for a statistical analysis
\citep{gou10}, as long as any observations taken in response to the detection of a
planetary signal are not included.

Most planetary signals are presently found by the high cadence surveys, and MOA-II, as
the first high cadence survey, began finding planetary signals in survey data back in 
2007 (or 2006 if non-alerted discoveries are included \cite{bennett12}). But, with only
a single high cadence observing site until 2010, and only partial coverage of the 
MOA high cadence fields afterwards, we have found that planetary signals identified in 
the MOA-II can often be characterized much better when follow-up data, as well as
OGLE and WISE survey,  are included in the light curve modeling. Therefore, in our analysis, 
we use only the MOA-II survey data for detection of potential planetary light curve
anomalies, but we use all available data for characterization of the binary light curve 
parameters.


\section{Event Selection}
\label{sec_eve_sel}

In 2007-2012, a total of 3300 microlensing 
alerts\footnote{https://it019909.massey.ac.nz/moa/} were issued by MOA. 
Most of the alerts are microlensing events, but some other types of transient events are also included.
To determine the exoplanet mass ratio function, it is necessary to construct a sample of well characterized 
single lens and planetary events with good coverage by MOA-II survey observations.
While it is possible to detect planets in events with a strong stellar binary signal
\citep{gou14,pol14}, the analysis of such events is substantially more complicated, so
we postpone such an effort to a later analysis. Therefore, events that are strong stellar
binary events or are obviously not due to microlensing are removed from our sample.
However, we do not rely upon simple light curve inspection to distinguish between stellar 
binary and planetary binary microlensing events. Instead, we have done a systematic
analysis of the binary microlensing events in our sample to determine the best fit
mass ratio, $q$. After confirming that an event passes our binary or planetary detection
criteria that the binary or planetary model should improve the fit $\chi^2$ by
$\Delta\chi^2 > 100$ using the data on the MOA alert page, we collect re-reduced data from 
all the observatories that have observed each event. This data is not used for determining
which events pass our selection criteria, but it is used to find the best fit models that
are used for event classification.

The modeling with these expanded data sets was performed with a grid search using the image-centered
ray-shooting method \citep{ben96,bennett-himag}.
As discussed by \citet{suz14}, to find the best fit model, we searched a wide range of the 
parameter space using a variation of the Markov Chain Monte Carlo (MCMC) algorithm \citep{ver03}, 
starting from 9680 grid points with fixed $q$, $s$ and $\alpha$ 
($s$: the projected separation normalized to the angular Einstein radius, $\alpha$: 
the angle between the source trajectory and the binary lens axis). Then, we searched $\chi^{2}$ minima 
of the best 100 models in order of $\chi^{2}$ in the first grid search with letting the each parameter free.
We also included the microlensing parallax effect \citep[e.g.]{mur11} to the model in this systematic analysis.
678 events were analyzed with this modeling effort. Independent modeling of all
unpublished planetary events and possible planetary events was also performed
using the initial condition grid method of \citet{bennett-himag}.

Following \citet{bon04}, we define planetary events to be ones with
a best fit mass ratio of $q < 0.03$. We also check for degenerate solutions that have best fit
$\chi^2$ values that differ by $\Delta\chi^2 < 25$ (after the error rescaling as discussed
below), and in these cases, we include 
all the models as possible models for the event. While a threshold of $\Delta\chi^2 = 10$
would be sufficient for Gaussian statistics, since it would imply a probability of $< 1\,$\%,
we use a higher threshold in order to enable the possibility of testing our results for
sensitivity to non-Gaussian effects.
We consider all possible solutions in
Bayesian analysis to determine the exoplanet mass ratio function.
For any binary events that might be potentially planetary
events, we attempt to improve the MOA photometry with a variety of methods, in order to
get the best possible event classification, and as mentioned in the last section, we include
data from other groups. But, the anomaly detection is based on the data directly from the
MOA alert pages.

The systematic modeling has recovered all the previously known planetary events and
detected four new planetary events. One of these four events was previously published 
as a stellar binary event, but our systematic analysis showed that it is clearly a 
planetary event. The other three planetary events were newly discovered by the 
systematic modeling effort, and their planetary signals were missed in the years in
which they occurred. Our treatment of all these planetary events is discussed 
in Section \ref{sub_sec_pla_eve}.
We find 23 planetary events (including one ambiguous event) and 1451 single 
lens events in our 6-year MOA-II survey data. This compares to 6 planets (in 5 events) and
8 single lens events in the high magnification sample of \citet{gou10} and 3 planets
and 196 single lens events in 6 years of PLANET follow-up photometry of events alerted by 
OGLE. One planetary event, OGLE-2007-BLG-349/MOA-2007-BLG-379 is common
to all three analyses, but there is no other overlap in these planetary event samples.


Here, we briefly summarize our classification of alert events. We classify the events 
into six categories: single lens, planetary, stellar binary, ambiguous, 
cataclysmic variable (CV) stars, and unknown events.
Single lens events have simple light curves, known as \citet{pac86} curves, and we use this
simple light curve form to identify well characterized single lens events with the selection criteria
given in Table~\ref{tab:criteria} as described in the following subsection.
For planetary (stellar binary) events, following three conditions are required: the MOA light 
curves are better fit by a binary lens model than a single lens model by $\Delta \chi^{2} > 100$;
the best fit mass ratio is smaller (larger) than 0.03 using all available data including 
OGLE and other follow-up data, and the best fit model is better than stellar binary 
(planetary) solutions by $\Delta \chi^{2} > 25$.
If the $\chi^{2}$ difference between the best planetary and stellar binary solution is smaller
 than $\Delta \chi^{2} = 25$, then the event is classified as an ambiguous event.
The light curves of CVs are not well fit by both single lens and binary lens model, 
and their light curves are so characteristic that they can be classified by visual inspection.
The events classified as ``other" are ones that fail the selection criteria 
(described in the next subsection) due to such problems
as poor light curve coverage or large systematic errors in the photometry. 
We avoid possible misclassification by setting our $\chi^2$ threshold to be
$\Delta\chi^2 > 100$, which is much higher
than would be necessary if we were only concerned with Gaussian random noise.
There are at least two planetary candidates that have signals below this 
detection threshold, 
OGLE-2007-BLG-292/MOA-2007-BLG-190 and
OGLE-2012-BLG-0724/MOA-2012-BLG-323 \citep{hir16}.
For the purpose of this paper, these are treated as non-detections, and they
are the price we pay for setting a threshold high enough to avoid misclassifications.


\subsection{Selection Criteria for Single Lens Events}
\label{sub_sec_cri}

Once the planetary, stellar binary events and CVs are removed from the sample, the remaining
events are the events classified as single lens or ``other" events.
In order to consider only those events that allow the detection of planets with 
well-defined parameters, we apply a set of
cuts, listed in Table~\ref{tab:criteria} on the remaining events. These
cuts rely upon single lens microlensing fits to the online MOA-II data. In those cases
where a finite-source microlensing model provides a better fit than a point-source model,
we use the finite-source model for cuts given in Table~\ref{tab:criteria}. Note that none of
these criteria are related to the presence or absence of planetary light curve signals.

In order to search for planetary signals in the microlensing light curves, we require well 
constrained single lens microlensing fit parameters. Otherwise, we cannot tell what types 
of planets the light curve might be sensitive to. Cut-1 and Cut-2 impose these constraints
and also exclude events that may not be due to lensing by stellar or brown dwarf primaries.
The flux, f, in the MOA online data is the delta-flux, which is the measured flux in the
difference image, as computed by the difference image method \citep{bon01}. The flux error,
fe, is the error bar computed by the photometry code, so Cut-3 is a measure of the
S/N of the primary microlensing signal above the baseline brightness. Cut-4 requires
a sufficient number of data points in the magnified part of the light curve to detect
a planetary signal, and Cut-5 ensures that there is reasonable coverage and
a reasonable S/N on both the rising and falling sides of the light curve. We also want to avoid
microlensing events which reach their single lens peak outside of or too close to the
end of the season, and this is the rationale for Cut-6. Cut-7 consists of three conditions
that must be satisfied for the event to be accepted. All three conditions ensure that the
event does not have large gaps in data coverage around the peak. The events failing
Cut-7 also have very low planet detection efficiency, so this cut has little effect on 
the total detection efficiency.

The result of applying these cuts on our sample of 3300 microlensing events is
that 1451 events pass all the cuts. These 1451 events consist of 
226, 230, 247, 288, 184 and 276 events, respectively, in each year of the range 2007 -- 2012.
Since the selection criteria we use are not related to the presence or absence of planetary signals, 
the rejection of poorly sampled events induces no bias. Our selection criteria do influence the
properties of the distribution of stars that are searched for planetary signals, but this
distribution is already influenced by unavoidable factors, such as observing conditions.
Some readers might be surprised at that only half of the observed events 
passed the event selection criteria, but this ratio is comparable to the previous studies.
\citet{cas12} and \citet{shv15} used 45\% and 49\% of observed events for their analysis.
\citet{gou10} used only 4\% of the observed events due to the strict cuts they used.

\begin{figure}
\begin{center}
\plotone{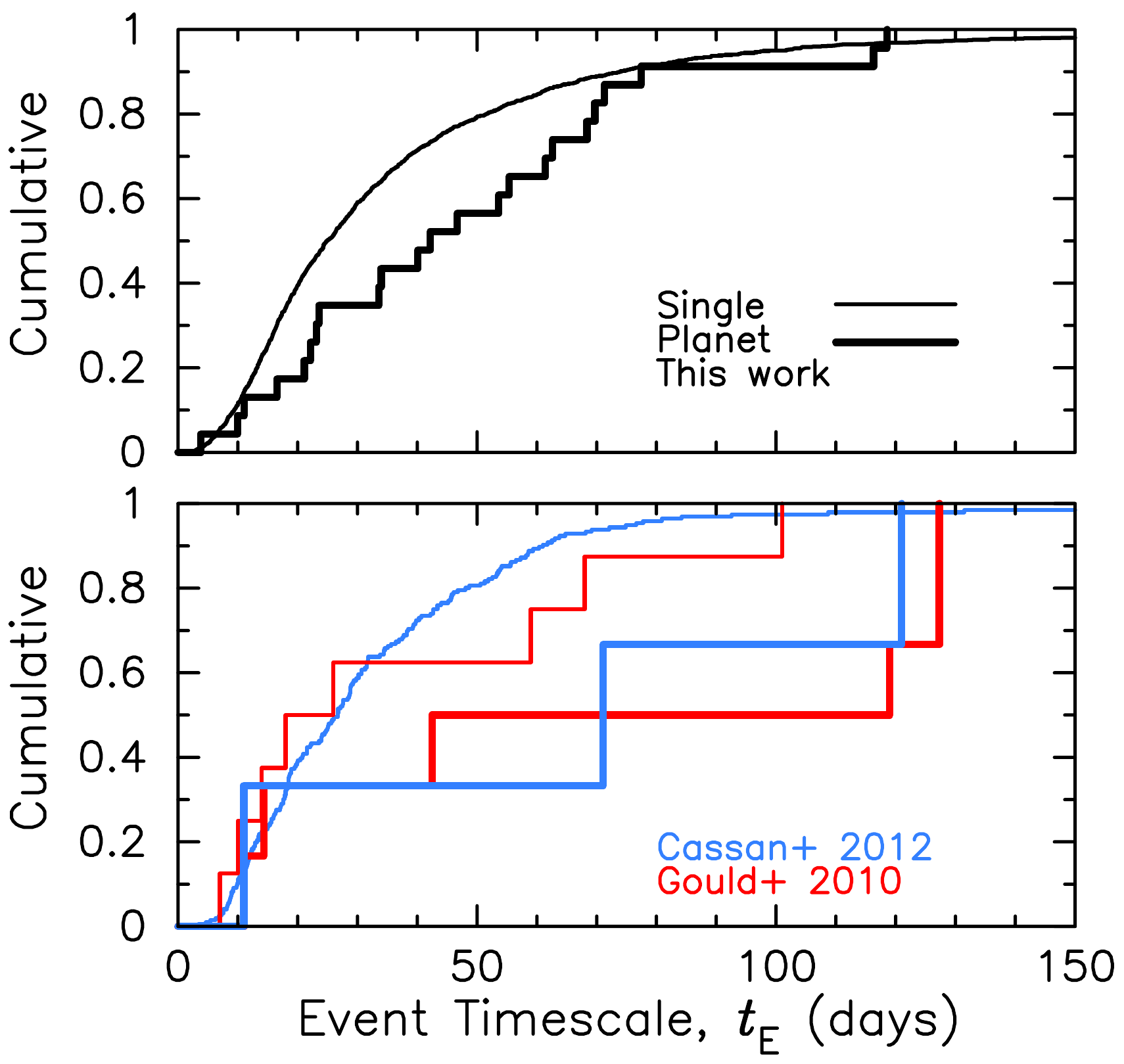}
\caption{
Cumulative distribution of the event time scale $t_{\rm E}$, for the 1474
single lens (thin lines) and planetary events (thick lines) that pass cuts 1--7.
The top panel compares our full sample to the planetary events in the sample.
The bottom panel shows the same $t_{\rm E}$ distributions for the previous statistical
samples of  \cite{gou10} and \cite{cas12}, respectively.
}
\label{fig:tE_dist}
\end{center}
\end{figure}

The cumulative $t_{\rm E}$ distribution for these 1451 events is shown in the top panel
of Fig.~\ref{fig:tE_dist}, along with the distribution of events with planets. As we show in
Section~\ref{subsec:fun_tE}, the difference between these distributions is largely, but
perhaps not entirely, a selection effect. The lower panel of Fig.~\ref{fig:tE_dist} shows
these same distributions for the $\mu$FUN \citep{gou10} and PLANET \citep{cas12}
statistical analyses in red and blue, respectively. Both show a similar bias toward planetary
signals in long events, while the $\mu$FUN distribution shows a slight bias toward
longer events in the full sample, including single lens events. The full $t_{\rm E}$ distribution 
for our sample is similar to that of the PLANET sample \citep{cas12}, which is 
only slightly skewed toward longer duration events, compared to our sample.


\subsection{Planetary Events}
\label{sub_sec_pla_eve}

The systematic modeling of all the anomalous events in our sample was conducted using 
all available data, including data from the OGLE survey and follow-up groups. This effort 
found four new planetary events, as well as the known planetary events that have either 
been published or are in preparation for publication. In order to determine the significance of
the planetary signals in the MOA data, all the planetary events were fit with both single lens
and planetary models using only the online MOA data. For possible planetary signals discovered
while they are in progress, MOA often takes additional data in order to better characterize
the parameters of candidate planetary microlensing events. We must be careful not to let
these additional data affect the planetary detection statistics, because data taken in response
to the detection of the planetary signal cannot be used to determine the detection efficiency.
So, the additional data taken in response to the anomaly detection are removed from the
fits used to determine which events pass the anomaly detection threshold. The detection
threshold fits use only the data taken at the standard MOA survey cadence. 
We calculate the fit $\chi^2$ for the best fit single lens and binary lens models, 
and we renormalize the error bars to set $\chi^2/{\rm d.o.f.} \equiv 1$ for the best fit model.
This process is necessary before using $\Delta\chi^2$ values as detection thresholds 
because the error bars provided by crowded field photometry codes are more accurate
measures of the relative rather than absolute photometric accuracy. This procedure
is commonly used in most of the previous microlensing analyses (e.g.\ Muraki et al.\ 2011).
Then we calculate the fit $\chi^2$ for the best fit single lens and 
planetary binary lens models again with these renormalized error bars.
If the difference between the renormalized 
$\Delta \chi^{2}$ values for the best fit single lens and the best fit planetary 
models is  $\Delta \chi^{2} > 100$, then the event passes our detection threshold.
There are four planetary events in the MOA data set that do not pass this detection
threshold cut because the planetary signals for these events were observed in other data
sets, so that a significant planetary signal is not seen in the MOA data. These events are
MOA-2007-BLG-400 \citep{don09mb07400}, MOA-2008-BLG-310 \citep{jan10}, 
MOA-2011-BLG-293 \citep{yee12}, and OGLE-2012-BLG-0724/MOA-2012-BLG-323 \citep{hir16}, 
and they are examples of lens systems with planets
that are not detected because of the low planet detection efficiency of the MOA
data for these events.

We require that our sample of planetary events should have that same good light curve 
coverage that we require for non planetary (single lens) events. For each possible planetary event, 
an artificial single lens light curve is generated using the parameters of the best fit 
planetary model and the gaussian noise based on the residuals of the best fit model.
Then, our selection criteria (see Section~\ref{sub_sec_cri} and  
Table~\ref{tab:criteria}) are applied to each artificial single lens light curve for each 
candidate planetary event.
We find that 23 events that are best fit by a binary lens model with a planetary 
mass ratio pass our cuts. However,
for one of these events, there is a competing model with a stellar binary mass ratio
that fits the data nearly as well. We classify this event as an ``ambiguous event\rlap",
which we treat in detail in Section~\ref{sub_sec_amb_eve}. This leaves the 22 clear 
planetary events, which we list in Table \ref{tab:planet}. It is the Einstein radius
crossing times, $t_{\rm E}$, of these 22 clear and 1 ambiguous planetary events
that are plotted in Figure \ref{fig:tE_dist}. The median $t_{\rm E}$ value for the 
planetary sample is $42\,$days, compared to $25\,$days for the full sample. Thus,
the typical planetary event is 1.7 times longer than the typical well sampled event.
This is primarily due to selection effects, as we discuss in
Section~\ref{subsec:fun_tE}.

Our sample of 22 clear planetary events includes four that were found for the first 
time by our systematic modeling effort. These events are
OGLE-2008-BLG-355/MOA-2008-BLG-288 \citep{kos14}, 
MOA-2008-BLG-379/OGLE-2008-BLG-570 \citep{suz14}, MOA-2010-BLG-353 \citep{rat15}, 
and MOA-2011-BLG-291 \citep{ben17}.
Here are short descriptions of these four events.
\begin{description}
\item[MOA-2008-BLG-288:] This event was initially announced as a stellar binary 
event in an analysis of OGLE-III archival data using only OGLE survey data \citep{jar10}.
But, our systematic modeling of all the anomalous events using all available data 
found that a planetary model is strongly favored for this event, by $\Delta\chi^2 = 810$.
The MOA data cover a caustic exit that is critical for the correct classification of this event.
\item[MOA-2008-BLG-379:] This is a high magnification event with a faint source star, 
$I_s = 21.3$. The faintness of the source meant that the strong anomaly at the peak 
dominated the apparent brightness variation, and it was not immediately recognized 
as a high magnification event. As a result, the planetary nature of the event was not
recognized  in real time. The planetary feature is characterized by the survey data alone.
\item[MOA-2010-BLG-353:] This event has a relatively weak anomaly caused by the
approach of the source to a cusp of a major image planetary caustic. The peak magnification 
is low and the source star is not so bright, and as a result, the planetary anomaly for this 
event has the smallest significance of all planetary signals that pass our threshold. The 
OGLE data does not cover the anomaly, but it helps to constrain the non-planetary 
light curve parameters.
\item[MOA-2011-BLG-291:] This is a high-magnification event that was alerted and observed by the
$\mu$FUN follow-up group, as well as the MOA, OGLE, and Wise survey telescopes. 
Nevertheless, the planetary signal occurs almost entirely in the MOA and OGLE data sets.
This event was identified as a possible planetary event shortly after the peak, but it was
not seriously investigated prior to our systematic analysis.
\end{description}

High magnification events ($A_{\rm max} \simgt 50$) account for 41\,\% 
of our clear planetary event sample, compared to two of the three (67\,\%) planetary 
events in the PLANET sample \citep{cas12} and all 5 of the planetary 
events (100\,\%) in the $\mu$FUN sample \citep{gou10} 
that are high magnification events. (But note that there is one high magnification event
common to all three samples.) It is natural that analyses based on follow-up data will be
dominated by high magnification events, because the planet detection efficiency per event
is much larger for high magnification events than for low magnification events \citep{griest98,rhi2000}.
For the microlensing follow-up results that have been published to date, 
the total detection efficiency in the sample (total sensitivity) has been dominated by 
the high magnification events.

High magnification events are relatively rare and the planetary caustics are usually larger
than the central caustics, so signals from planetary caustics in low-magnification events
are expected to dominate the planetary discoveries made in survey mode, and this
is what we find with our sample. Most high magnification planetary events suffer from 
close/wide separation degeneracy \citep{griest98}, because the shape of central caustic which
depends on the mass ratio, $q$ and the absolute value of the logarithm of the star--planet 
separation, $|\log \,s|$ \citep{chu05}. So, the central caustics are nearly identical
for separations of $s$ and $1/s$. There is a similar degeneracy with the location 
of the caustics for low-magnification events, but the shapes of the caustics for $s<1$ and
$s>1$ are very different. Thus, observations of low magnification events can generally
break this $s \leftrightarrow 1/s$ degeneracy, and the separation is uniquely determined.
This means that low-magnification planetary events are suitable for study of planet distribution as 
a function of separation beyond the snow line. More planetary signals in low 
magnification events will be expected in the future survey observation.

The planetary signals that we detect in microlensing light curves are due to the source star
crossing or approaching a caustic or cusp. The caustic crossing events are of special 
interest because caustic crossing features in the light curves allow us to measure the 
angular Einstein radius, $\theta_{\rm E}$, as mentioned in Section \ref{sec_intro}.
We find that $23\,$\% of our planetary events do not include measured caustic crossings. This
percentage is larger than the $\sim\!\!12 \%$ of published microlensing planet discoveries 
without measured caustic crossings, but it is substantially smaller than the 
$\sim\!\!55 \%$ of the planet detections predicted by simulation \citep{zhu14} of a three-site 
high cadence survey like the Korean Microlensing Telescope Network \citep[KMTNet]{kim16kmtnet}.
This simulation makes a number of optimistic assumptions, but it illustrates the point that
many of the planetary signals detected by higher sensitivity surveys will be in non-caustic
crossing signals from higher mass ratio planets ($q \simgt 10^{-3}$). This is borne out by
the non-caustic-crossing planetary deviations in our sample.
But MOA-2012-BLG-527 \citep{kos16} is different from the other non-caustic-crossing 
planetary signals, because the source star crossed the demagnified region between the 
central caustic and two minor image, triangular planetary caustics. This region provides
a strong negative perturbation that makes it easier to detect low mass ratio 
($q \sim 10^{-4}$) planets.

In these non-caustic-crossing events, the detection of finite source effects is difficult. So, the
source radius crossing time is not well constrained for
MOA-2010-BLG-353 (with a $2.3\,\sigma$ signal), MOA-2011-BLG-322 \citep{shv14}, and 
MOA-2012-BLG-527. However, the mass ratios are measured precisely in these events.
This means that we must assume the source radius crossing times when calculating the
detection efficiency for these events, but this assumption has no significant effect on
the results we present in this paper.

In our planet sample, five planetary events have not been published yet: 
MOA-2010-BLG-117 \citep{ben16bin}, MOA-2011-BLG-291(which is described 
above), MOA-2012-BLG-006/OGLE-2012-BLG-0022 \citep{pol16}, MOA-2012-BLG-355 
(OGLE et al. in prep), and MOA-2012-BLG-505 \citep{nag16}.
Here, we briefly describe each event.
\begin{description}

\item[MOA-2010-BLG-117:] This is a low magnification event with a binary source star whose 
trajectories crossed almost perpendicular to the minor image caustics. The best fit planet 
model that assumes a single source star failed to explain the demagnified part between the 
two triangular caustics. The observed flux at the trough was brighter than the model. This
is well fit by adding a bright companion to the source, and the binary source model fits the
data much better than a single source, triple lens model, which could also explain the 
excess flux at the time of the trough. This event was also
observed by the OGLE-IV survey, but OGLE did not announce any alerts in the first
year of the OGLE-IV survey.

\item[MOA-2012-BLG-006:] This event was detected as an apparent short time scale 
event, but a few month later the magnification rose again with a much longer duration,
but a lower amplitude. Light curve modeling \citep{pol16} indicates 
that the initial magnification was caused by a planet located at $s \sim 4$, which is the largest 
separation among the discovered microlensing planets except for OGLE-2008-BLG-092LAc
\citep{pol14}.

\item[MOA-2012-BLG-355:] This is a low magnification planetary event, but 
an RR Lyrae star is located very close to this event on the sky. It contaminates the light curve, 
which slowed the identification of the planetary signal by a couple days. The proximity to
a known variable prevented its detection in the OGLE EWS system. 
We removed the contamination from the MOA and OGLE data, and the best fit model was
 found as shown in Table \ref{tab:planet}.

\item[MOA-2012-BLG-505:] This event is a short, high magnification event that was only
observed by MOA. An anomaly lasting only a few hours over the light curve peak was well 
covered due to MOA's high cadence observing strategy. As often occurs with high 
magnification events, there is a close-wide degeneracy, but it does not imply a large
uncertainty in $s$ because both values are close to 1.

\end{description}

False positives are generally not a problem for planetary microlensing signals, unlike
the cases of planetary transit \citep{mul16} and RV \citep{rob14} signals.
The only types of false positive that have been identified for planetary microlensing
events are binary source stars
\citep{gaudi98} and variable source stars microlensed at high magnification \citep{gou13}.
Both of these types of false positives are intrinsically rare because a binary companion
to the source must be microlensed at high magnification to mimic a planetary microlensing
event and because variable source stars are themselves intrinsically rare. Furthermore,
most planetary microlensing light curves have shapes that are not compatible with
false positives. In our sample, only MOA-2010-BLG-353 has a light curve that might be
compatible with a false positive. It is a low magnification event, so source variability with
an amplitude large enough to mimic the microlensing signal would easily be seen in
the baseline data, but a binary source might be a possibility. We fit a binary source
model to this event and find a best fit $\chi^2$ that it is worse than the best fit 
planetary model by $\Delta\chi^2 = 20.3$. The error bars during the
MOA-2010-BLG-353 planetary signal are $\simgt 4$\%, which is well above the
usual level of the correlated errors that are sometimes seen in microlensing light
curves. Thus, it is safe to exclude the binary source model for this event.

Errors in light curve modeling might also be considered to be false positives, but these
are not a problem for our analysis because of our systematic modeling procedure. All anomalous
events have been systematically modeled by at least two independent investigations, and
potential planetary events have been modeled by at least three independent efforts.
The recently published OGLE event, OGLE-2013-BLG-0723 provides a good example of
the effectiveness of the MOA light curve modeling program. The initial paper
\citep{udalski_ob140723} claimed the discovery of a Venus-mass planet orbiting
a brown dwarf in a binary system. After this paper was published, the photometric
data for this event was obtained from the authors, and one of us (DPB) began
systematic modeling of the event, using the procedure described in \citet{bennett-himag}.
This systematic modeling effort quickly revealed the correct model, which includes
only a stellar binary with no planet and a much smaller microlensing parallax signal. 
The authors of the original paper were contacted, and this resulted in a paper with
the correct model \citep{han_ob130723}.

Photometry for published events should be available permanently on the NASA 
Exoplanet Archive (http://exoplanetarchive.ipac.caltech.edu/), while the data for 
all planetary events in this sample,
including those that have yet to be published can be obtained directly from the
first or second author.

\subsection{Ambiguous Events}
\label{sub_sec_amb_eve}

We define the mass ratio threshold $q < 0.03$ as the upper limit for planetary events.
Binary lens systems with larger $q$ values are considered to be 
stellar binary events. However, for a number of events, there are several models
that can fit the data. In most cases, these degenerate models are caused by a well known
degeneracy between different planetary models, such as the $s \leftrightarrow 1/s$
degeneracy discussed in Section~\ref{sub_sec_pla_eve}, but there can also be
degeneracies between planetary and stellar binaries, as discussed by \citet{cho12b}.
In fact, one of the events analyzed by \citet{cho12b}, OGLE-2011-BLG-0950/MOA-2011-BLG-336, 
is in our sample. We cannot arbitrarily ignore or accept such an event, because either
choice would bias our measurement of the exoplanet mass ratio function. So, we
include both possibilities in a Bayesian analysis, as discussed below.

Our systematic analysis of all binary events turned up a number of other potentially ambiguous
events, but further light curve analysis with optimized MOA photometry revealed that most
of these were unambiguous stellar binary events or contaminated by systematic photometry
errors. Events were classified as ambiguous based upon the $\chi^2$ values of the 
best fit stellar binary and planetary models ($\chi^2_{\rm SB}$ and $ \chi^2_{\rm PL}$
respectively). These $\chi^2$ values are calculated with renormalized error bars to give
$\chi^2/{\rm d.o.f.} = 1$, and we classify an event as ambiguous if  
$ | \chi^2_{\rm SB} - \chi^2_{\rm PL} | < 25$. With this definition, the only ambiguous event
is the one identified by  \citet{cho12b}, OGLE-2011-BLG-0950/MOA-2011-BLG-336. 
The other possible ambiguous events did not pass our event selection cut.
\citet{cho12b} found that the planetary model was favored by $\Delta \chi^2 = 105$, but considered the
event to be ambiguous because of evidence for systematic photometry errors. Our 
optimized photometry has removed much of these systematic photometry problems, but
this also reduced the $\chi^2$ difference between the best stellar binary and planetary
models to $\Delta \chi^2 = 19.80$, favoring the planetary
model. This is the difference between the best fit stellar binary and planetary models, 
but there are two models in each category, due to the close-wide degeneracy for
high magnification events. The wide model is favored in each case, as can be seen
in Table~\ref{tab:ambiguous}, which lists the light curve model parameters and fit
$\chi^2$ values for each of the four (approximately) degenerate for this event.

The probability that this ambiguous event is due to a planetary system rather than a 
stellar binary depends both on the $\chi^2$ values for the respective fits and the prior
probabilities of lens systems with the parameters of the best fit models. We define the
relative fit probability of the stellar binary model to be $e^{-\Delta \chi^{2}/T}$ where
$T = 2$ is the correct value if the observational uncertainties follow an uncorrelated 
Gaussian distribution. When $T=2$, this is equivalent to the likelihood ratio
that is sometimes used for similar model comparisons \citep{boz16}.
We can adjust $T$ in order to test the effects of non-Gaussian
uncertainties.

In order to determine the prior probability for the stellar binary
model, we need to estimate the mass of the primary star, and we do this with 
a Bayesian analysis using constraints from the observed $t_{\rm E}$ values. 
We conduct the analysis with the same manner as used in the previous planetary events \citep[e.g.]{jp06}. 
The Galactic model of \citet{hg03} was used for the prior.
This yields a median primary star mass of $\sim 0.6 M_{\odot}$. The period and mass ratio
distribution for stellar binaries have been determined for solar type stars by \citet{rag10}
and for low-mass stars by \citet{del04}. Since the likely primary mass for the
OGLE-2011-BLG-0950/MOA-2011-BLG-336 lens system is intermediate between
the 
solar type and low-mass stars, we use the average value of 
the probability for solar type and low mass stars.  For solar type stars, the stellar 
multiplicity is $\sim \! 41 \%$, while for low-mass stars it is $\sim \!26 \%$. 
For the probability of a planetary companion with the parameters indicated by
each fit, we use our mass ratio function. The error bars in  Table~\ref{tab:ambiguous}
are also important, as they serve to determine the range in the mass ratio, $q$,
and separation, $s$, that are consistent with the light curve data, and these factors
tend to enhance the probability of the stellar binary interpretation, because the stellar
binary models are consistent with a wider range of separations and mass ratios
than the planetary models.
The priors that we derived from these considerations were 
0.196294 and 0.072682 for the wide and close planetary models and
0.530297 and 0.200727 for the wide and close stellar binary models, respectively.

In order to obtain the final probabilities for the different models we must multiply
these priors by the relative probabilities, $e^{-\Delta \chi^{2}/T}$, for the different
models. With the standard, $T=2$, value for uncorrelated Gaussian uncertainties,
we find final probabilities of 0.825276 and 0.174628 for the wide and close
planetary models and 0.000078 and 0.000017 for the wide and close stellar
binary models, respectively. These values are used for our final results, but we have also
considered the possibility of strong correlations or non-Gaussianity in the 
photometric uncertainties. If we increase $T$ to $T=20$, ten times the Gaussian random value,
then the relative probabilities become 0.375144 and 0.131346 for 
the wide and close planetary models and 0.363331 and 0.130179 for the wide and 
close stellar binary models, respectively. So, this would imply roughly equal probabilities
for the planetary and stellar binary models. Fortunately, even such an extreme
assumption of correlated or non-Gaussian errors, does not have a significant
effect on our final results.

\section{Planet Detection Efficiency}
\label{sec_est_de}

The detection efficiency for a planetary signal in a microlensing event is denoted by
$\epsilon(\log s, \log q)$.
This is the probability that a planet with its mass ratio, 
$q$ at the projected separation (in units of the Einstein radius), $s$, is detected,
assuming a random orientation for the planetary system.
Following \citet{rhi2000}, we use the best fit single lens model for each event, while
for events with planetary signals, we generate an artificial single lens light curve
using the parameters of the best fit planetary model.
We inject planetary signals with a complete range of parameters ($\log s, \log q, \alpha$) 
covering all of the potentially detectable signals to each single lens light curve.
The generated light curves have same observation times as actual MOA survey data. 
The error bars in each data point are re-normalized to be 
$\chi^{2}/{\rm d.o.f.} \equiv 1$ for the best single lens or planetary model as 
discussed in Section \ref{sub_sec_pla_eve}.
Note that any additional MOA observations taken in
response to a planetary signal are removed at this stage, 
as described in Section \ref{sub_sec_pla_eve}.
For each of these generated light curves, we search for the best single lens model, we then calculate
$\Delta \chi^{2} = \chi^{2}_{\rm single} - \chi^{2}_{\rm planetary}$ to determine if the
planetary parameters give a large enough signal to pass our detection threshold.
We do not need to calculate $\chi^{2}_{\rm planetary}$ because it remans the same 
as the original $\chi^{2}$ for single lens fit without binary signature added.
We use 360 grid points to cover the angle between the source trajectory and 
lens axis, $\alpha$, and define $\epsilon(\log s, \log q)$ as the fraction 
of grid points for which $\Delta \chi^{2}(\log s, \log q, \alpha) > \chi^{2}_{\rm thresh}$.
Of course, we must use the same $\chi^2_{\rm thresh}$ as we used as the
detection threshold for the planetary signals, discussed in Section~\ref{sub_sec_pla_eve}.
So, we set $\chi_{\rm thresh}^{2} = 100$. Microlensing is insensitive to inner planets 
with $s < 0.1$ and has very low sensitivity for outer planets with $s > 10$, except in
events where the microlensing signal of the host star is not directly detected
\citep{bennett12}. So, for the purpose of the detection efficiency calculations, 
we sample $s$ with 40 logarithmically spaced bins spanning the range $0.1 \leq s \leq 10$. 
Since we know that the detection efficiency depends smoothly on $q$,
we use 9 logarithmically spaced bins covering the range $ -5 \leq \log  q \leq -1.5$.
Thus, the total number of artificial light curves simulated for one event is 129600.
For the handful of events sensitive to planets with smaller mass ratios, we extend
the grid down to $\log  q = -6$.

Formally, if we assume that the photometric errors follow an uncorrelated gaussian 
distribution, then our detection threshold of $\chi_{\rm thresh}^{2} = 100$ corresponds
to a false alarm probability of $\sim 2\times 10^{-23}$. However, we do not believe that
our error bars follow an uncorrelated gaussian distribution. Also, we want to ensure that
the planetary signal is detected with enough S/N to provide reliable determinations
of the planetary light curve parameters, $q$ and $s$. Note that the analysis of high
magnification events by \citet{gou10} used a much higher threshold, $\chi^{2}_{\rm thres} = 500$.
However, this threshold was applied to data that was re-processed after a planetary 
signal was suspected. Since some of the original data was reduced by DoPHOT
\citep{dophot} instead of more precise difference imaging photometry, the reprocessed 
data is likely to have a substantially larger
$\Delta\chi^2$ than the original data. Also, much of the data in the \citet{gou10} 
sample was taken with unfiltered telescopes. This means that these data are vulnerable to
systematic photometry errors due to the color dependence of atmospheric transparency
\citep{don09ob05071,bennett-ogle109}. Finally, the high magnification events are
often observed continuously all night, and this means that the effect of correlated 
photometric errors are much more significant than with the MOA data, with a 
cadence of at most one observation every 15 minutes. For these reasons, \citet{gou10}
thought that their higher threshold was reasonable for their analysis.

\subsection{Source Star Radius}
\label{sub_sec_ss_radius}

The size of a source star is important for the detection of planets, because 
planetary light curve features are broadened and depressed with increasing source star size. 
The finite source effect is characterized by the source radius crossing time, $t_*$, which is the time
needed for the lens to cross the angular source star radius, $\theta_{\ast}$.
A larger source star has more chance to cross caustics, but it usually weakens the 
deviation from the single lens light curve. For the detection of planets with giant
source stars and $ q \simlt 10^{-4} $ or main sequence sources and $ q \simlt 10^{-5} $
the source star size is critical because the planet signals can be smoothed out 
by finite source effects \citep{ben96}. Thus, finite source effects can have important
effects on the detection efficiency, so we need to estimate $t_*$ for each event.

For the planetary event, we use the $t_*$ value from each paper or the best fit planetary model.
Although the $t_*$ is not well constrained for MOA-2010-BLG-353, MOA-2011-BLG-322 and 
MOA-2012-BLG-527, we used the $t_*$ of the best fit model.
For the high-magnification events with lens stars passing over source stars, MOA-2007-BLG-176, 
MOA-2007-BLG-233/OGLE-2007-BLG-302, OGLE-2008-BLG-290/MOA-2008-BLG-241, 
MOA-2010-BLG-436, MOA-2011-BLG-093, and OGLE-2011-BLG-1101/MOA-2011-BLG-325, 
we use the $t_*$ value for each event from \citet{cho12a}.
Also, we use the $t_*$ of \cite{bat09}, \cite{gou09}, \cite{yee09}, \cite{bac12hi}, \cite{yee13},
 \cite{gou13}, \cite{fre15} and \cite{par14} for
OGLE-2007-BLG-050/MOA-2007-BLG-103, OGLE-2007-BLG-224/MOA-2007-BLG-163, 
OGLE-2008-BLG-279/MOA-2008-BLG-225, MOA-2009-BLG-411, MOA-2010-BLG-311, 
MOA-2010-BLG-523, MOA-2011-BLG-274 and OGLE-2012-BLG-0455/MOA-2012-BLG-206, respectively.
Note that MOA-2009-BLG-411 and MOA-2012-BLG-206 are stellar binary events 
with a brown dwarf companion or possible planetary events, but the anomaly 
signals are found in the follow-up data, instead of MOA data. So, they are treated 
as single lens events in this analysis.
In addition to the above events, we use the best fit $t_*$ value estimated with all 
available data for the following events because they are probably showing the 
finite source effect: MOA-2008-BLG-383, MOA-2011-BLG-393, 
MOA-2012-BLG-278/OGLE-2012-BLG-1430, OGLE-2012-BLG-0617/MOA-2012-BLG-285, 
OGLE-2012-BLG-1098/MOA-2012-BLG472, and OGLE-2012-BLG-1274/MOA-2012-BLG-538.

For the other single lens events, we need to estimate $t_*$ by estimating both $\theta_{\ast}$ 
and the relative lens-source proper motion, $\mu_{\rm rel}$, since 
$t_*$ = $\theta_{\ast}/\mu_{\rm rel}$.
In most planetary microlensing events, the light curves are observed with different filters to 
estimate $\theta_{\ast}$ using the empirical relation of the $\theta_{\ast}$, color and 
magnitude of the source star \citep{kervella_dwarf,kerfou08,boy14}. For the 2007-2011
observing seasons, we do not have source color information for the single lens events,
because the only filter used for regular MOA observations was the MOA-Red wide-band filter.
But, the source star magnitude in the MOA-Red passband, $R_{\rm M,S}$, is determined 
by the light curve model for each event. The absolute magnitude of the source star, 
$M_{I, \rm S}$, can then be estimated with the formula 
$M_{I, \rm S} = R_{\rm M,S} - (R_{\rm M,RC} - M_{I, \rm RC})$, where $R_{\rm M,RC}$ is 
the apparent magnitude of red clump giant (RCG) centroid in the MOA-Red passband, and 
$M_{I, \rm RC}$ is the absolute magnitude of RCG centroid in the $I$-band.
$R_{\rm M,RC}$ is estimated from the color-magnitude diagram (CMD) of the reference images 
in the $R_{\rm M}$ and $V$-bands in each subframe (1k $\times$ 1k section of each CCD).
We assume $M_{I, \rm RC} = -0.12$ following \cite{nat13}.
Using the calculated $M_{I, \rm S}$ value with the 10 Gyr isochrone model from 
the PARSEC isochrones, version 1.2S \citep{bre12,che14,che15,tan14}, 
the source star radius can be estimated. The source star distance
depends on Galactic longitude, and we use the values from \citep[Table 1]{nat13}
to determine $\theta_\ast$.

We need to estimate the lens-source relative proper motion, $\mu_{\rm rel}$ to determine $t_*$.
$\mu_{\rm rel}$ ranges over about an order of magnitude, with 
$1\,{\rm mas/yr} \simlt \mu_{\rm rel}\simlt 10\, {\rm mas/yr} $.
As discussed in \citep{ben14}, for events with both the source and lens located in the bulge, the 
typical value is $<\mu_{\rm rel}> \simeq 5.6\,\rm mas/yr$.
For bulge-source disk-lens events, we have $<\mu_{\rm rel}> \simeq 8.2\,\rm mas/yr$.
We assume the typical value for our microlensing events is $<\mu_{\rm rel}> \simeq 5.6\,\rm mas/yr$,
because typical lens star is located at the bulge. Of course, this is an 
underestimate of $\mu_{\rm rel}$ for cases when the lens star is located at the disk, which means it
overestimates $t_*$ for these events. But these overestimates are by only about  $50\,$\%,
and this has a very small effect on our final results. 

\begin{figure}
\begin{center}
\epsscale{0.9}
\plotone{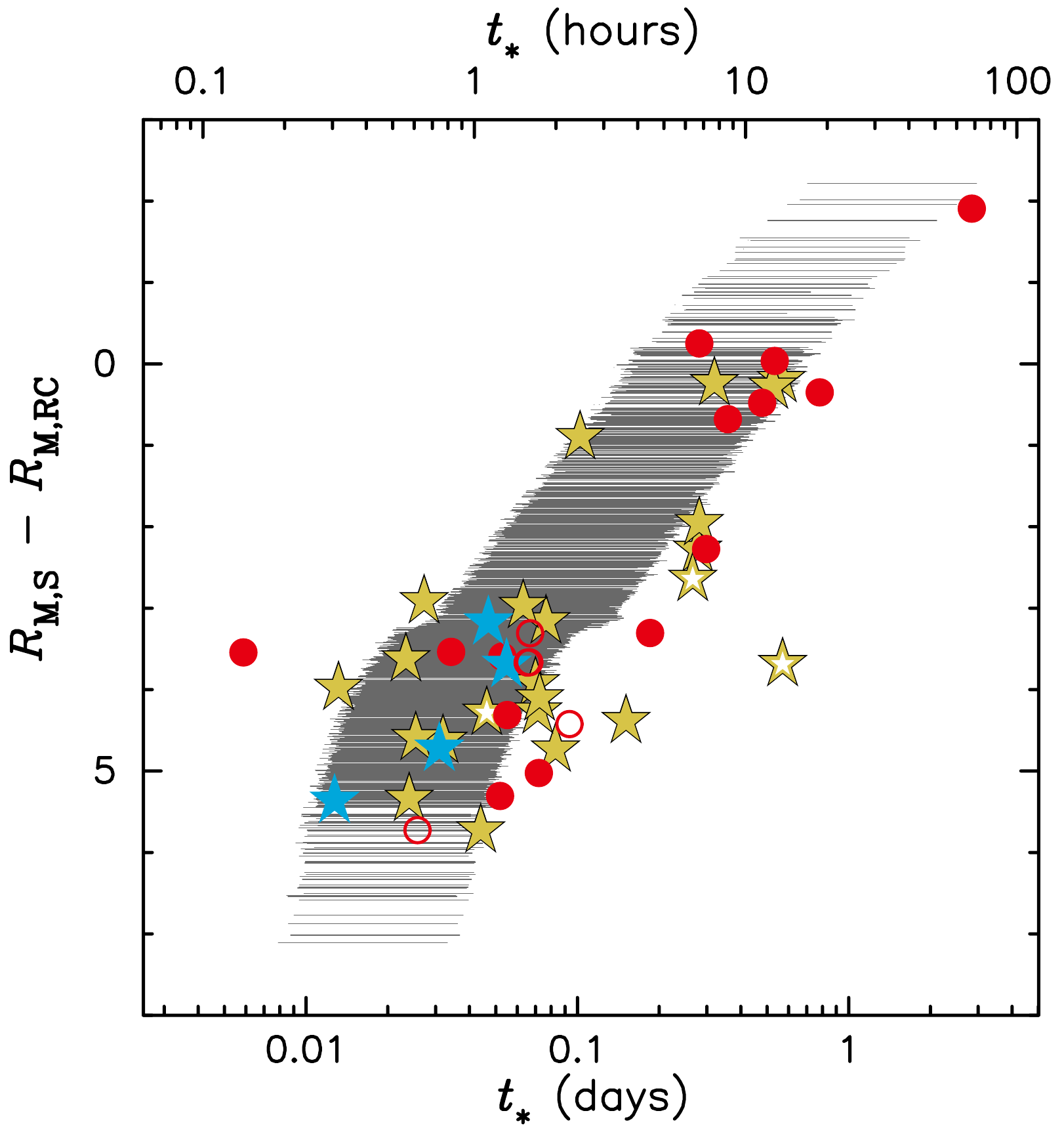}
\caption{
Source radius crossing time $t_\ast$ vs. the relative magnitude of source stars to red clump giant (RCG) centroid in MOA-Red filter.
The gray lines indicate single lens events which need the estimated $t_\ast$ to compute the detection efficiency.
The length of the gray lines indicate $1\,\sigma$ uncertainties of estimated $t_\ast$.
The gold filled stars show the planetary events with measured $t_\ast$ in our sample, whereas the gold open stars are the planetary events with poorly determined $t_\ast$.
The blue stars are also planetary events with measured $t_\ast$, but 
they are not included in our planet sample
because they do not show enough anomaly signals in the MOA data.
The red circles indicate the events with a clear finite source effect. Most of them are high magnified single lens events, but some events are stellar binary events that do not show the anomaly signals in the MOA data. The red filled and open circles are published and not yet published events, respectively.
}
\label{fig:rho}
\end{center}
\end{figure}

To check whether our results are sensitive to our treatment of finite source effects,
we consider the uncertainty in $t_*$, which is a quadrature sum of the uncertainties 
in $\theta_*$ and $\mu_{\rm rel}$. We assume a $30\,$\% error for the estimated 
$\theta_*$ values. Our planetary event sample yields an average relative proper motion 
of $\mu_{\rm rel} = 5.74_{-2.84}^{+2.94}$ mas/yr, and we assume that the single lens events
have the same $\mu_{\rm rel}$ distribution as the planetary events. In
Figure \ref{fig:rho}, we compare the estimated uncertainties in $t_\ast$ to the measured 
$t_\ast$ values  for planetary events and finite source events.
The dimensionless source star size is given by $\rho = \theta_*/\theta_{\rm E}=t_* / t_{\rm E}$, 
and its uncertainty, $\sigma_{\rho}$, for each event is determined from the
uncertainties in $\theta_*, \mu_{\rm rel}$ and $t_{\rm E}$. 
The detection efficiency will be computed in the next section with three different values
of the dimensionless source size: our estimated value, $\rho$, as well as 
$\rho_{\rm min}$and $\rho_{\rm max}$, where $\rho_{\rm min}  = \rho - \sigma_{\rho}$ 
and $\rho_{\rm max} = \rho + \sigma_{\rho}$.

\subsection{Detection Efficiency as a Function of Separation and Mass Ratio}

\begin{figure}
\begin{center}
\plotone{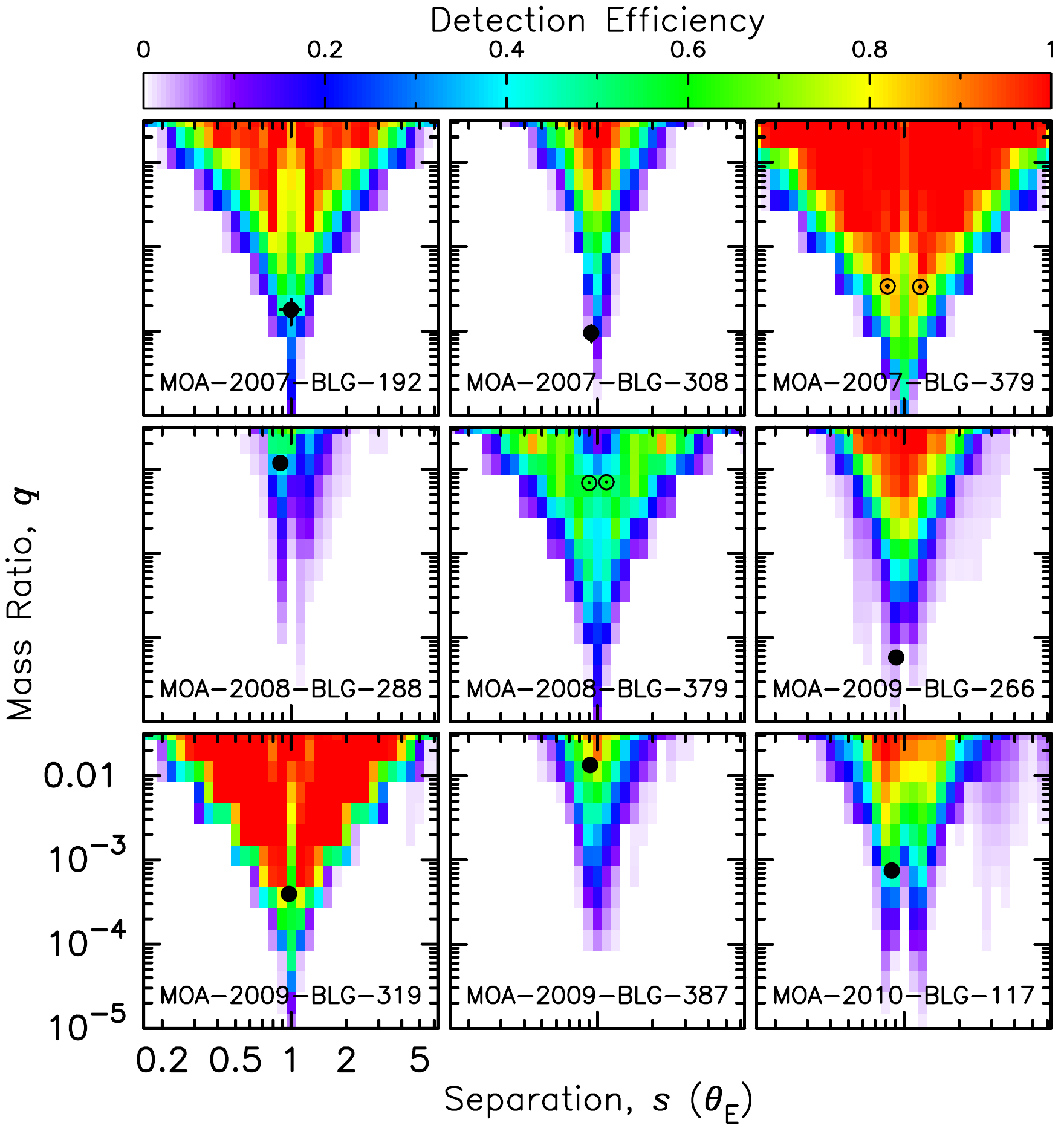}
\caption{
Detection efficiencies $\epsilon (\log s, \log q)$ for the planetary events.
The filled circles indicate the locations of the detected planets and
the open circles are used for the events with the close/wide separation degeneracy.
}
\label{fig:dep1}
\end{center}
\end{figure}

\begin{figure}
\begin{center}
\plotone{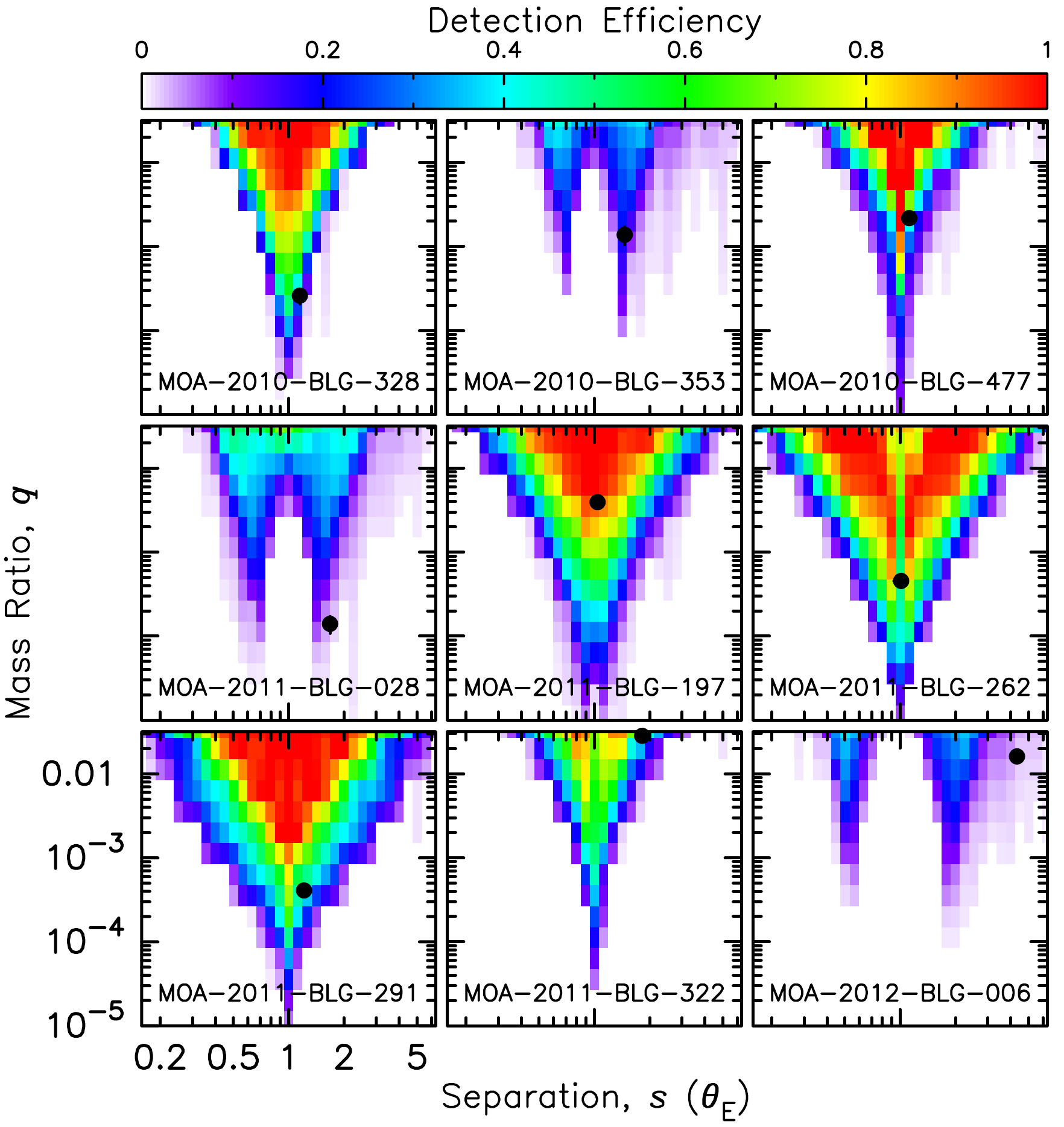}
\caption{
Same as Figure \ref{fig:dep1}.
}
\label{fig:dep2}
\end{center}
\end{figure}

\begin{figure}
\begin{center}
\plotone{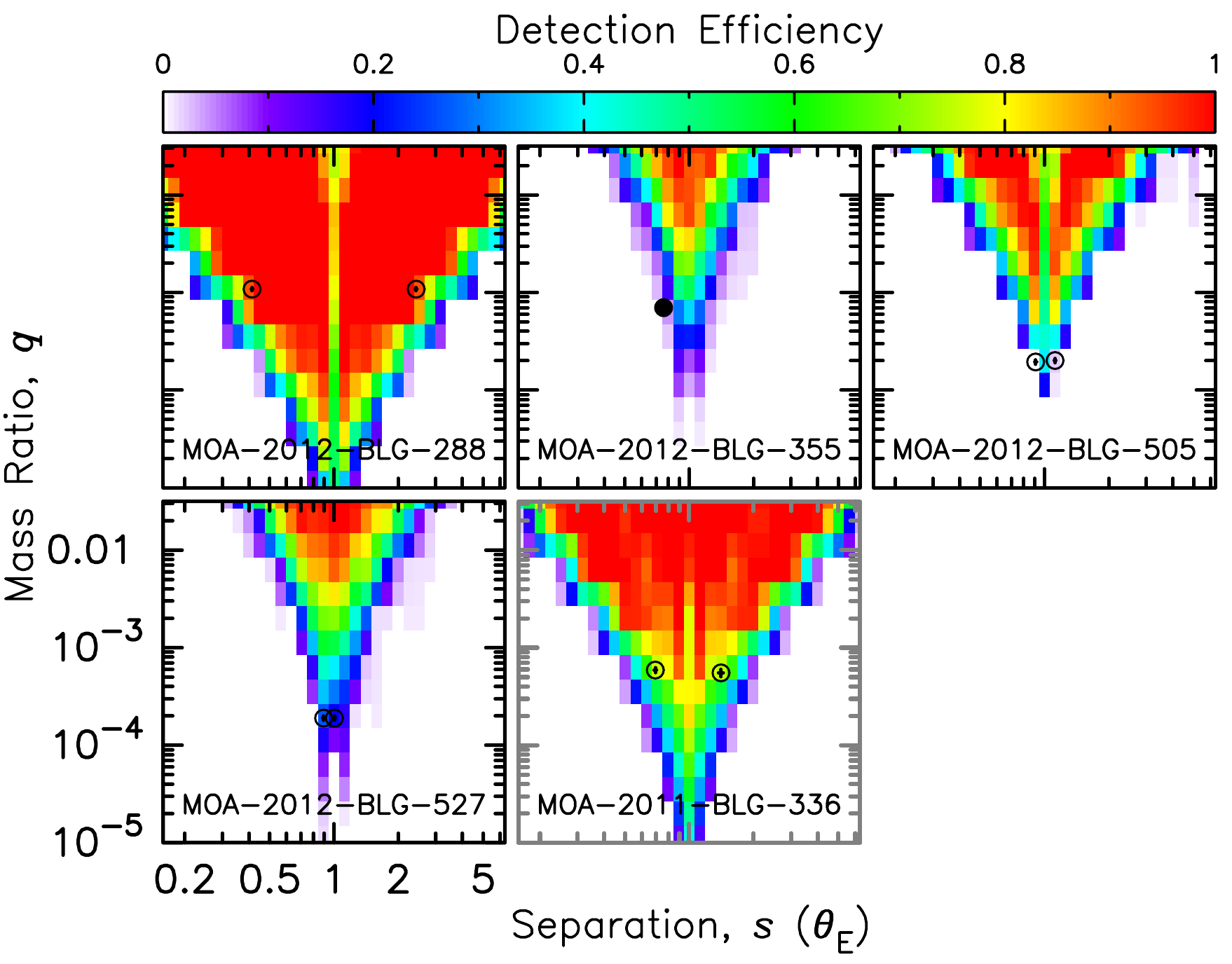}
\caption{
Same as Figure \ref{fig:dep1}, but the last panel (with the gray boxes) is the
detection efficiencies for the ambiguous event, where the planetary model is used for 
the calculation of detection efficiency.
}
\label{fig:dep3}
\end{center}
\end{figure}

We have computed the detection efficiency $\epsilon(\log s, \log q)$ for each of the 
1474 events that passed the event selection criteria discussed in Section \ref{sec_eve_sel}.
Figures~\ref{fig:dep1} - \ref{fig:dep3} show the detection efficiency for the 
planetary and ambiguous events. Typically, each figure has roughly triangular
sensitivity contours as first shown by \citet{rhi2000}, and subsequently shown
by a number of subsequent studies \citep{don06,yee09,bat09,gou10,cho12a}.
But, low magnification events tend to show a double peak toward lower mass ratios
at $s>1$ and $s<1$ and reduced detection efficiency at the Einstein ring radius ($s = 1$).
These features were also seen in \cite{ben96} and \cite{gau02}.

\begin{figure}
\begin{center}
\plotone{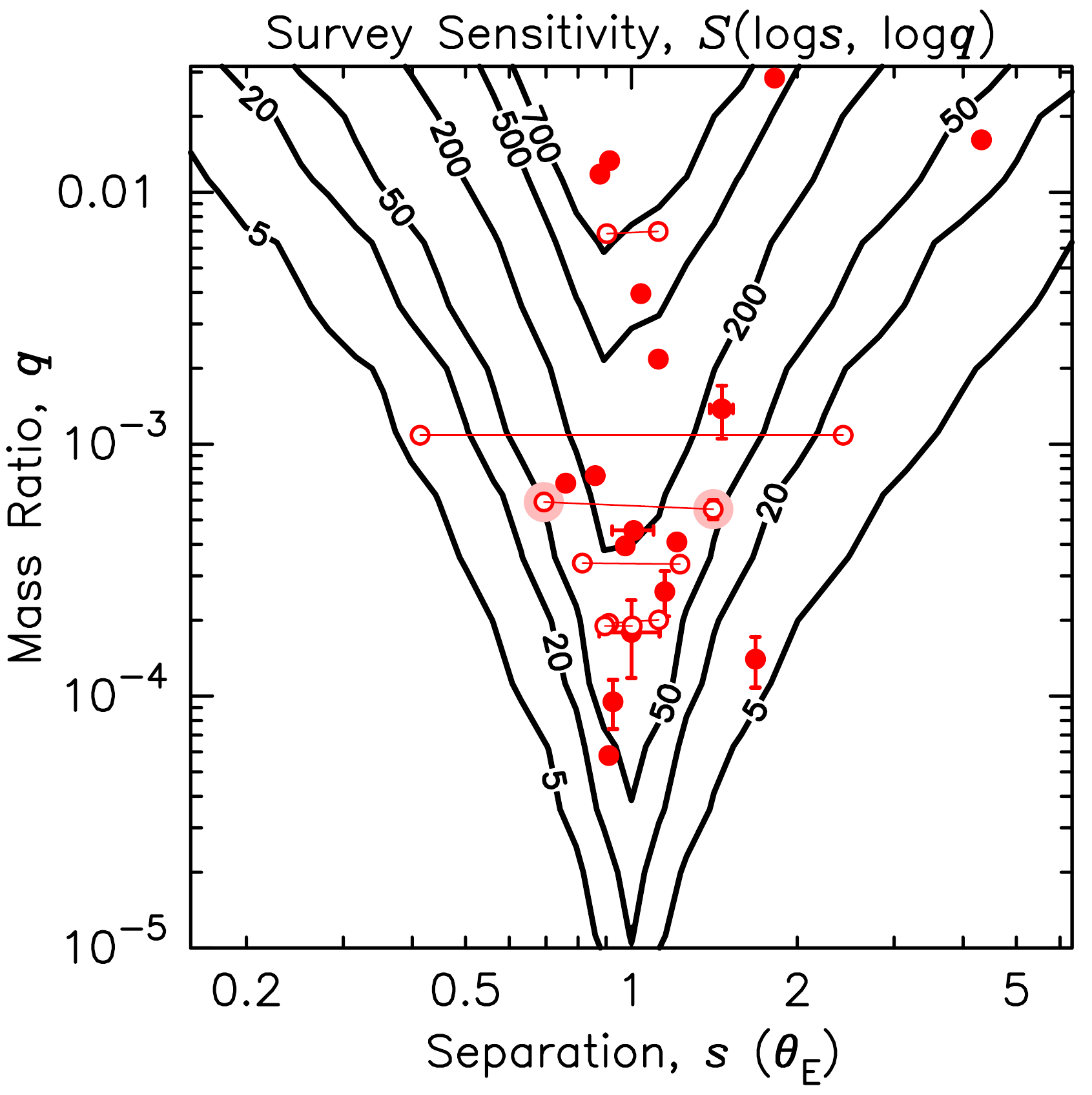}
\caption{
Sum of the detection efficiencies, or survey sensitivity $S(\log s, \log q)$, plotted as a function 
of $\log s$ and $\log q$. The contours indicate the number of planets detected if each lens star
has a planet at the specified $s$ and $q$ values.
The average detection efficiency is obtained by dividing the survey sensitivity by the total
number of events, 1474. The 23 planets in the sample are plotted with the filled and open circles
for the planets without and with the separation degeneracy, respectively.
The close and wide solutions for each separation degeneracy event are connected with a line, and
the ambiguous event is highlighted with light red halos.
}
\label{fig:sensitivity_s_q}
\end{center}
\end{figure}

When we sum the detection efficiency for all the events in our analysis, we obtain
the survey sensitivity, 
\begin{equation}
S(\log s, \log q) \equiv \sum \epsilon(\log s, \log q)\ ,
\label{eq:surv_sens}
\end{equation}
which is plotted in Figure \ref{fig:sensitivity_s_q}.
The survey sensitivity $S(\log s, \log q)$ is the number of planet detections that we expect 
with our 1474 events sample
\footnote{ The survey sensitivity and detection efficiency can be obtained by
contacting the first or second author.},
if we assume the uniform planet distribution in $\log s$ and $\log q$.
If our survey was dominated by well-sampled, high-magnification events, then the
survey sensitivity contours would be nearly symmetric in $\log  s$ \citep{gou10,cho12a}.
However, most of the sensitivity in our sample comes from more modest magnification
events. This leads to higher sensitivity for planets with $s > 1$ than for planets with $s < 1$.
This is due to the fact that outer planets perturb the higher magnification ``major" image
instead of the lower magnification ``minor" image. For a single lens event with magnification
$A$, the major image has a magnification of $(A+1)/2$ while the minor image has a
a magnification of $(A-1)/2$, so when the magnification approaches $A = 1$, we can
only expect to detect planets that perturb the major image.  
As shown in Figure \ref{fig:sensitivity_s_q}, most of the planets discovered in our data
have separations close to the Einstein radius, $s \sim 1$, and the distribution is pretty
flat in $\log q$ with slightly more planets with $10^{-4}< q < 10^{-3}$ than between
$10^{-3}< q < 10^{-2}$. Since we are less sensitive to planets with small $q$ values,
this implies that lower mass planets are more common than more massive planets,
at least down to $q \sim 10^{-4}$.

\begin{figure}
\begin{center}
\plotone{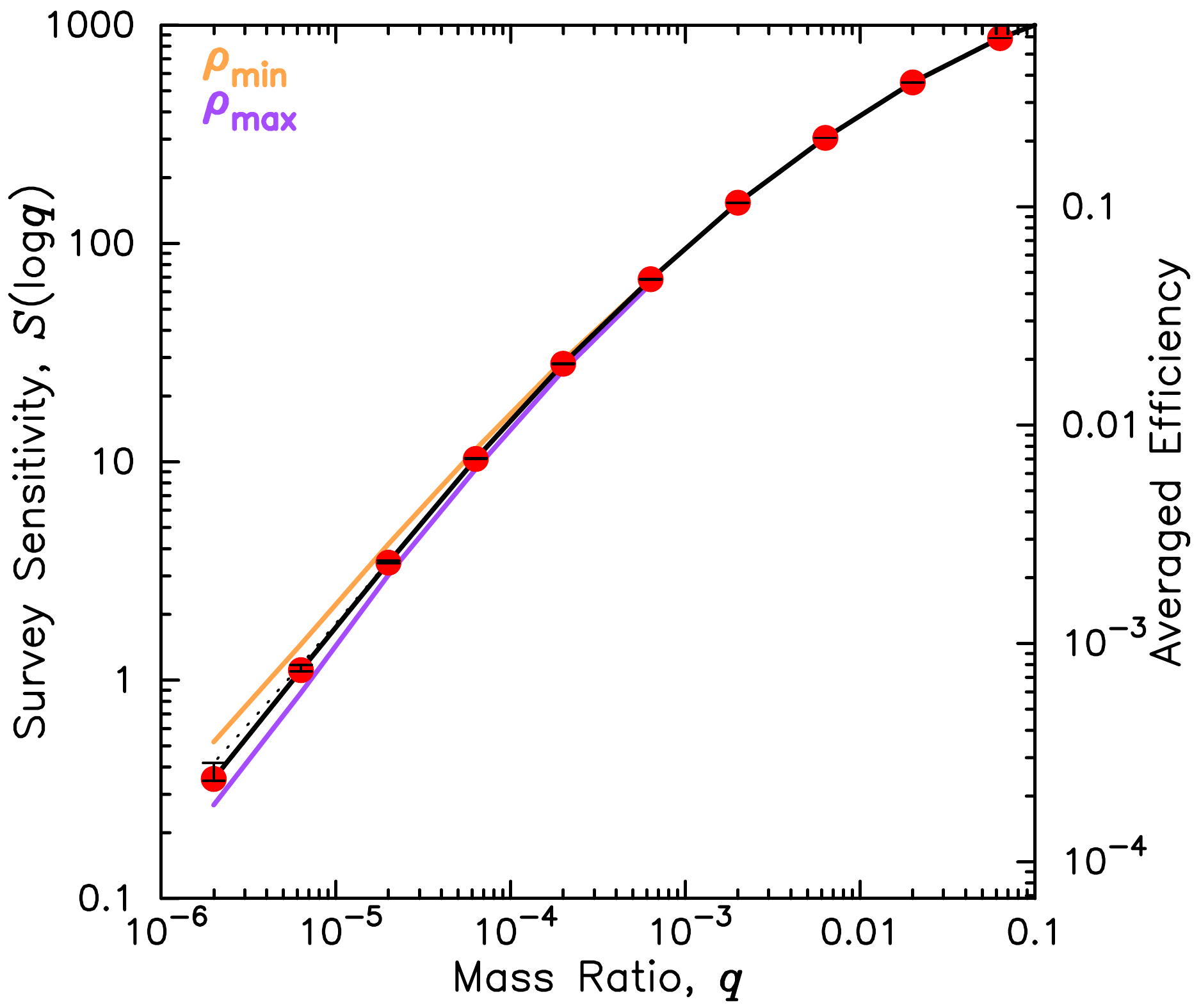}
\caption{
Survey sensitivity as a function of $\log q$, derived by integrating $S(\log s,\log q)$ over $\log s$ 
over the range $0.1 \leq s \leq 10$, plotted along with the left Y-axis scale.
The red points show the survey sensitivity, and the black solid lines are the linearly interpolated 
sensitivity between the calculated grid points.
The statistical uncertainties are plotted with error bars and dotted lines.
Also, the sensitivity estimated with the $1\, \sigma$ lower and upper values of the finite source size, $\rho_{\rm min}$ and $\rho_{\rm max}$ are plotted in the orange and purple lines, respectively. The uncertainty of the finite source size is negligible above $q \sim 10^{-3}$, so they are not plotted.
If the survey sensitivity is divided by the total number of events 1474, then we obtain averaged detection efficiency, which is shown on the right Y-axis.
}
\label{fig:sensitivity_q}
\end{center}
\end{figure}

We can also define the survey sensitivity as a function of $\log q$, by
integrating $S(\log s, \log q)$ over $\log s$,
\begin{equation} 
S(\log q) = \int_{\log s_{-}}^{\log s_{+}} S(\log s, \log q) d\,\log s,
\label{eq:surv_sens_logq}
\end{equation}
where $s_{-} \leq s \leq s_{+}$ is $0.1 \leq s \leq 10$.
Figure \ref{fig:sensitivity_q} shows the survey sensitivity, $S(\log q)$, where we have linearly 
interpolated the sensitivity between each calculated grid point.
The statistical uncertainties in the sensitivity are negligible.
The survey sensitivities estimated using the $\rho_{\rm min}$ and $\rho_{\rm max}$, 
which are $1\,\sigma$ lower and upper values of $\rho$, are also plotted in Figure \ref{fig:sensitivity_q}.
We use the two-dimensional $S(\log s, \log q)$ to determine the exoplanet mass 
ratio function in the following section, instead of the integrated, 1-dimensional
survey sensitivity, $S(\log q)$.

\section{Planet Frequency}
\label{sec_pla_fre}

The planet-star mass ratio, $q$, is measured in virtually all planetary microlensing events, 
while the masses can be sometimes determined by subtle microlensing parallax effects or 
follow-up observations with high angular resolution, but the masses of the lens stars and planets have
not been measured for a large fraction of events. So, it is natural to determine the exoplanet
frequency as a function of $q$. Similarly, the projected separation, $s$, is readily determined in units of
the angular Einstein radius, and it can generally only be determined in physical units for the same
events that allow the determination of lens star and planet masses. Our choice of parameters
differs from that of \citet{cas12}, who attempted to determine the exoplanet frequency as a 
function of planet mass without the use of any mass measurements. 
So, the masses could only be estimated with
a Bayesian analysis that assumed that the planet frequency does not depend on the
host star mass. 
We feel that simply reporting the mass ratio function is a better representation
of the constraints of the exoplanet population from microlensing observations.

\subsection{As a Function of Separation and Mass Ratio}
\label{subsec:fun_q_sep}

\begin{figure}
\begin{center}
\plotone{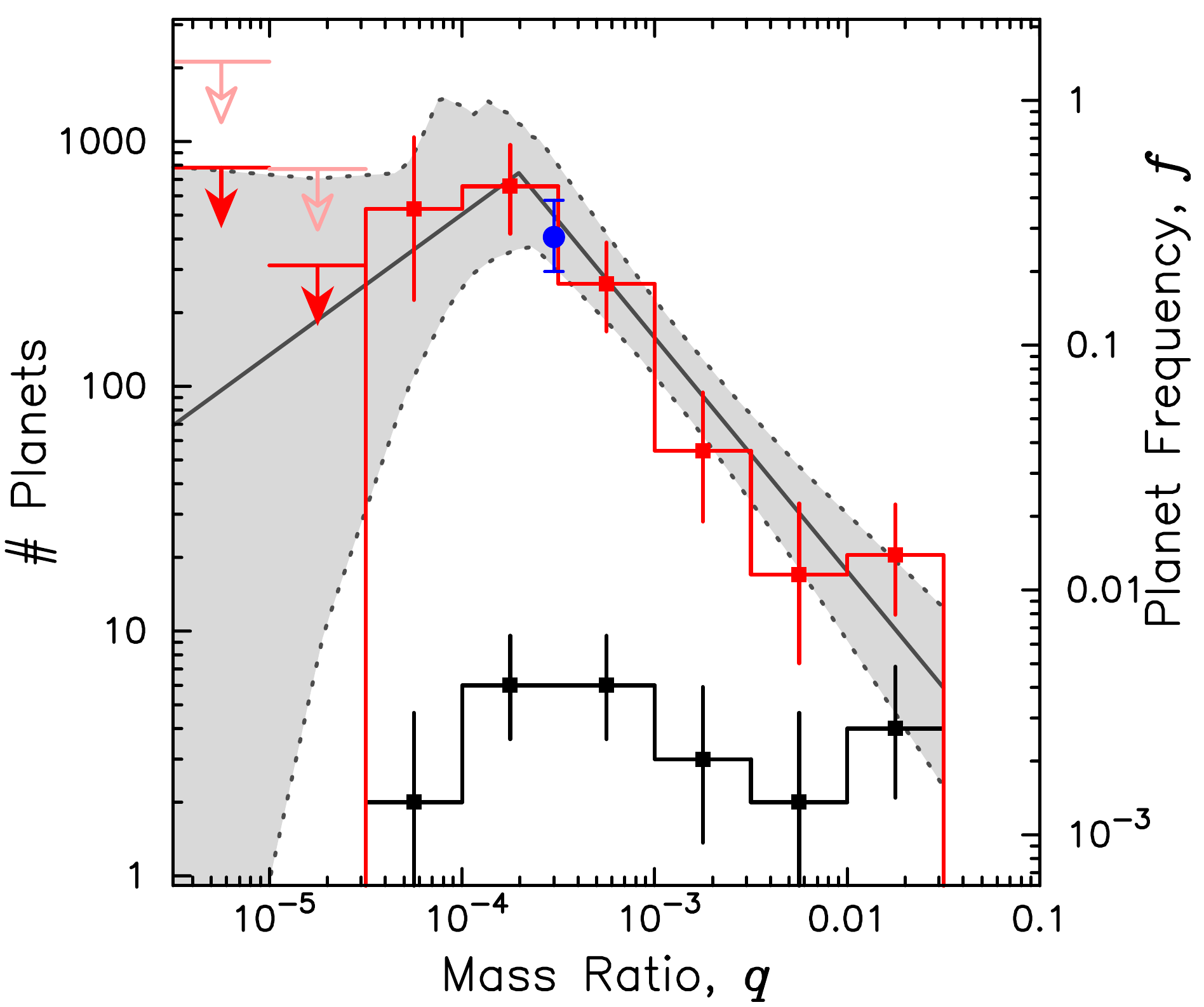}
\caption{
Mass-ratio distribution of the planets in this analysis.
The black histogram shows the number of the detected planets including the one ambiguous event.
The number of detected planets corrected by the survey sensitivity in Figure \ref{fig:sensitivity_q} is 
plotted in red, with 1-$\sigma$  error bars. 
No planets were found in the bins with $\log q < -4.5$, so the upper limits with 1-$\sigma$ and 2-$\sigma$
confidence level are plotted with red and pink arrows, respectively.
The best fit broken power-law mass ratio function,
evaluated at $s=1$, is shown in black, and the gray
shaded region covers the area mapped out by the broken power-law mass ratio function models
that are consistent with the data at the 68\% confidence level. The blue point and error bar
indicate the 
median
and 1-$\sigma$ range of the exoplanet mass function at $q = 3\times 10^{-4}$.
}
\label{fig:q_func}
\end{center}
\end{figure}

Figure~\ref{fig:q_func} shows histograms of the the distribution of planet mass ratios 
in this analysis. The black histogram is the number of detected planets as a function
of mass ratio, and the red histogram includes a detection efficiency correction that considers a poisson probability. 
So, the red histogram represents our estimate of the true planetary
distribution, with 1 and 2-$\sigma$ upper limits in the non detection mass ratio bins. 
Note that these histograms are only for visualization. The unbinned data is used 
for the calculations that determine the mass ratio function.
There are several features that are apparent from this distribution. 
First, the detected distribution is pretty flat down to $q \sim 10^{-4}$ or $5\times 10^{-5}$.
This implies that the frequency of planets increases at small $q$ with a slope
that nearly cancels the slope of the survey sensitivity function shown in Figure~\ref{fig:sensitivity_q}.
However, it is also clear that this trend does not continue much below $q \sim 10^{-4}$.
An extrapolation of the slope of the sensitivity-corrected mass ratio function at $q > 10^{-4}$
toward $q< 10^{-4}$ passes well above the 2-$\sigma$ upper limits for the bins with
$q < 3\times 10^{-5}$. This indicates that the power-law mass ratio functions used by
\citet{gou10} and \citet{cas12} will not be able to explain our data.

Another shortcoming of the previous microlensing studies of \citet{gou10} and \citet{cas12} 
is that they did not have enough events to measure
the separation dependence of the mass ratio function. This difficulty is exacerbated by
the close/wide degeneracy for the high magnification events that dominate the 
\citet{gou10} and \cite{cas12} samples. In particular, only 2 of the 6 planets in the \citet{gou10}
sample and 3 of the 8 planets in the \cite{cas12} had the close/wide degeneracy resolved.
In contrast, the close-wide degeneracy is resolved in 17 of the 23 events in our sample
due to the fact that most of the planetary signals in our sample were discovered through 
the planetary-caustic channel as expected for a high cadence survey.
Therefore, our sample can be used to measure the exoplanet separation distribution.

The considerations above lead us to consider an exoplanet mass ratio function consisting
of a power law in projected separation, $s$, and a broken power-law in the mass ratio, $q$,
\begin{eqnarray}
f(q,s;A,q_{\rm br},n,p,m) &\equiv& \frac{d^{2} N_{\rm pl}}{d\,\log q\ d\,\log s} \nonumber \\
&=& A \left[\left(\frac{q}{q_{\rm br}} \right)^{n}\Theta(q-q_{\rm br}) +  \left(\frac{q}{q_{\rm br}} \right)^{p}\Theta(q_{\rm br}-q)\right] s^{m},
\label{eq:expl_mrf}
\end{eqnarray}
where $N_{\rm pl}$ is an average number of planets around a lens star, 
$n$ and $p$ are the power law indices for the mass ratio above and below the 
break in the mass ratio function at $q = q_{\rm br}$. The power law index for the separation is
denoted by $m$, and $\Theta$ is the (Heavyside) step function. 

For events with multiple degenerate models that fit the data, we weight the different models
with $e^{-\Delta \chi^{2}/2}$, where $\Delta \chi^{2}$ is the $\chi^{2}$ difference between 
the model in question and the best fit model. Mostly, these degenerate models are close and
wide models, but we also use a similar procedure for our ambiguous planetary event
OGLE-2011-BLG-0950/MOA-2011-BLG-336, as discussed in Section~\ref{sub_sec_amb_eve}.

We use a likelihood analysis to determine the exoplanet mass ratio function parameters,
$A$, $q_{\rm br}$, $n$, $p$, and $m$. It is straight forward to derive the likelihood function
from Poisson statistics \citep{alc96,gou10}, and it is given by
\begin{eqnarray}
\mathcal{L}(A, q_{\rm br}, n, p, m) &= & e^{-N_{\rm exp}} \prod _{i}^{N_{\rm obs}} f(s_{i},q_{i};A,q_{\rm br},n,p,m) S(\log s_{i},\log q_{i}) \nonumber \\
N_{\rm exp} &=& \int ds \int  dq\ f(s,q;A,q_{\rm br},n,p,m)S(\log s,\log q) \ ,
\label{eq:q-s}
\end{eqnarray}
where $N_{\rm exp}$ is the number of planets detected for a mass ratio function model, 
$f(q,s;A,q_{\rm br},n,p,m)$, with
the measured survey sensitivity, $S(\log s_{i},\log q_{i})$, defined in Equation~(\ref{eq:surv_sens}).
The parameters $q_i$ and $s_i$ refer to the mass ratio and separation (in Einstein radius units)
for each of the $N_{\rm obs}$ observed planets. The uncertainties in $q_i$ and $s_i$ are
included as a two dimensional Gaussian distribution. (This modified the model parameters
by $< 0.1$-$\sigma$ compared to a 
calculation that ignored the error bars.)
The best fit exoplanet mass ratio function parameters are the ones that maximize 
the likelihood function, and they are shown in Table~\ref{tab:expl_bmrf}.

In order to derive confidence intervals for the exoplanet mass ratio function parameters
using a Bayesian approach, we must chose ``prior distributions" of these parameters,
which are equivalent to the integration measure that we use to integrate over the
the likelihood function, $\mathcal{L}(A, q_{\rm br}, n, p, m)$. We chose uniform priors
for $n$ and $m$, while for $A$ and $q_{\rm br}$ we chose priors uniform in
$\log A$ and $\log q_{\rm br}$. The power index, $p$, for planets with small mass ratios,
$q < q_{\rm br}$, presents a more difficult choice of a prior. We only have a small number
of planet detections for $q \simlt 10^{-4}$, so mass ratio functions with $q_{\rm br}$ close
to the $q$ value of the lowest mass ratio planet in our sample are not strongly 
excluded by our data set. Such models are consistent with very large $p$ values,
which would imply that planets with $q \simlt 10^{-5}$ are extremely rare. But, 
we know that such planets exist from {\it Kepler} and solar system examples, and we
also expect that it is relatively  common for planets to change orbits due to gravitational
interactions with other bodies. 
Similarly, values of $p \ll -1$ are also unlikely due to planetary scattering considerations. 
Planet distributions determined from the {\it Kepler} survey indicate that small planets are quite
common in shorter period orbits \citep{how12,don13,for14}, so there should be some 
small planets in wider orbits, if only due to planet-planet scattering.
Thus, it is sensible to chose a prior that disfavors priors with $|p| \gg1$.
Hence, we chose a prior distribution that is uniform in hyperbolic arcsine.
This prior reduces $|p|$ by about a factor of two, and changes the other parameters
by much less than their error bars.
The uncertainties in $q$ and $s$ for each planetary event are included as the noise 
model for this analysis, as discussed above.

\begin{figure}
\begin{center}
\plotone{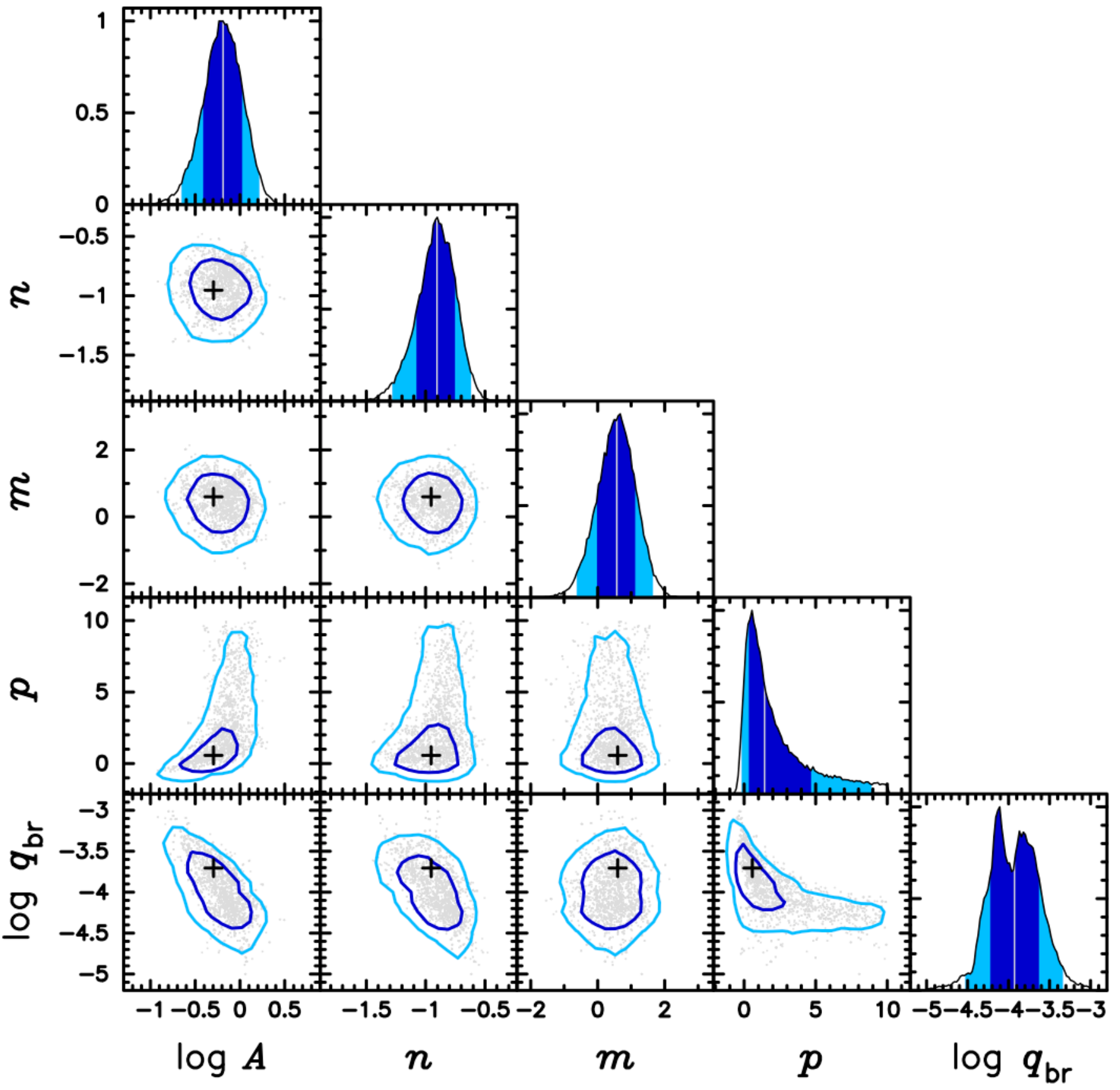}
\caption{
Likelihood contours for the broken double power law function, $f =  A \left[\left(\frac{q}{q_{\rm br}} \right)^{n}\Theta(q-q_{\rm br}) +  \left(\frac{q}{q_{\rm br}} \right)^{p}\Theta(q_{\rm br}-q)\right] \left(\frac{s}{s_{0}}\right)^{m}$
for the MOA sample.
The contours indicate the 68\% and 95\% confidence intervals, and the crosses indicate the
best fit parameters (which are also listed in Table~\ref{tab:expl_bmrf}).
}
\label{fig:mrf_contMOA}
\end{center}
\end{figure}

The maximization of the likelihood function was conducted with the
Markov Chain Monte Carlo (MCMC) algorithm 
to determine the best fit parameters and uncertainties for each parameter 
using the Affine Invariant MCMC Ensemble sampler of \citet{emcee}. The best
fit values are shown as crosses in Figure~\ref{fig:mrf_contMOA}, and they are also
listed in Table~\ref{tab:expl_bmrf}. Figure~\ref{fig:mrf_contMOA} also shows 
scatter-plots and 68\% and 95\% confidence levels of the MCMC results, and 
Table~\ref{tab:expl_mrf} shows the median value and 68\% confidence level
for each parameter. Although the uncertainty of the power-law index, $p$ is large, 
the power law break at $q\sim10^{-4}$ is clearly detected.
The best fit model with a broken power law function is favored over the single power law
model by $\Delta \chi^{2} = 11.6$ with two additional model parameters. The $\chi^2$ distribution
with two degrees of freedom indicates that the single power-law model can be excluded
with $p$-value of 0.0030. However, this use of the $p$-value to estimate the evidence for
the additional model parameters has been criticized as being insufficiently conservative.
\citet{trotta08} argue that it is better to use the Bayesian evidence or Bayes factor, which 
includes some accounting of the Occam's razor preference for a simpler model. Following
\citet{trotta08}, we can estimate the Bayes factor as $\mathcal{B} = -1/(\mathcal{P} e \ln \mathcal{P})$, 
where $\mathcal{P}$ is $p$-value, 
and this yields
a Bayes factor of $\mathcal{B} = 21$, implying that the broken power law model is favored by a factor of 
21 over the single power law model.

Figure~\ref{fig:q_func} shows the behavior of
the exoplanet mass ratio function at $s = 1$.
The uncertainties of the estimated finite source sizes for events without
finite source  are also statistically negligible. 
If we use the survey sensitivity estimated with $\rho_{\rm min}$ values, the changes of the best fit model 
parameters are less than $0.1\, \sigma$. If we use 
the $\rho_{\rm max}$ values, which reduce the sensitivity for low-mass-ratio planets, 
the normalization, $A$, changes by $0.1\,\sigma$.

It is clear from Figures~\ref{fig:q_func} and \ref{fig:mrf_contMOA} that the amplitude
and slope of the mass ratio function above the break at $q = q_{\rm br}$ are pretty
well constrained. But, the value of $q_{\rm br}$ and the amplitude $A$
of the mass ratio (measured at $q_{\rm br}$) are not determined so precisely. 
The power-law index, $p$, of the mass ratio function below the break is even more
poorly determined.
This is due to the small number of planets found with $q < 10^{-4}$ and the lack
of planets with $q < 3\times 10^{-5}$. The choice to normalize the mass ratio function
at $q = q_{\rm br}$ allows us to write the mass ratio function, Equation~(\ref{eq:expl_mrf})
in a more compact way, but it also increases the apparent uncertainty in the 
mass ratio function at larger $q$ values. Therefore, we introduce the parameter
$B = f(q=3\times 10^{-4},s=1)$, shown in Table~\ref{tab:expl_mrf} and 
Figure \ref{fig:q_func} with the blue point, to indicate the
uncertainty in the normalization at the mass ratios where microlens planets are more typically found.

\subsection{As a Function of Event Time Scale}
\label{subsec:fun_tE}

Microlensing has the potential to determine the exoplanet mass function beyond the
snow line as a function of host mass and galactocentric distance, but this requires additional
information such as microlensing parallax \citep{ben08,gau08,bennett-ogle109,mur11,sko14}
or detection of the lens star \citep{ben06,ben07,ben15hst,don09ob05071,bat14,bat15}. 
While such measurements exist for some of the planets in our sample, our analysis
does not use this information. Nevertheless, we can still explore the dependence
of the exoplanet mass ratio function on the one microlensing parameter, $t_{\rm E}$, that 
depends on the mass and galactocentric distance of the host star. 

Shorter $t_{\rm E}$ events can be caused by low-mass lens stars or a combination of 
a high lens-source relative velocity and a small lens-source distance. These later conditions
hold for lens systems in the Galactic bulge, and a systematic trend of planet properties
with $t_{\rm E}$ would likely indicate a trend in planet properties with host star mass
or galactocentric distance, or some combination of the two effects. (The mass and distance
estimates alluded to in the previous paragraph are needed to break this degeneracy
between lens mass and distance.)

\begin{figure}
\begin{center}
\plottwo{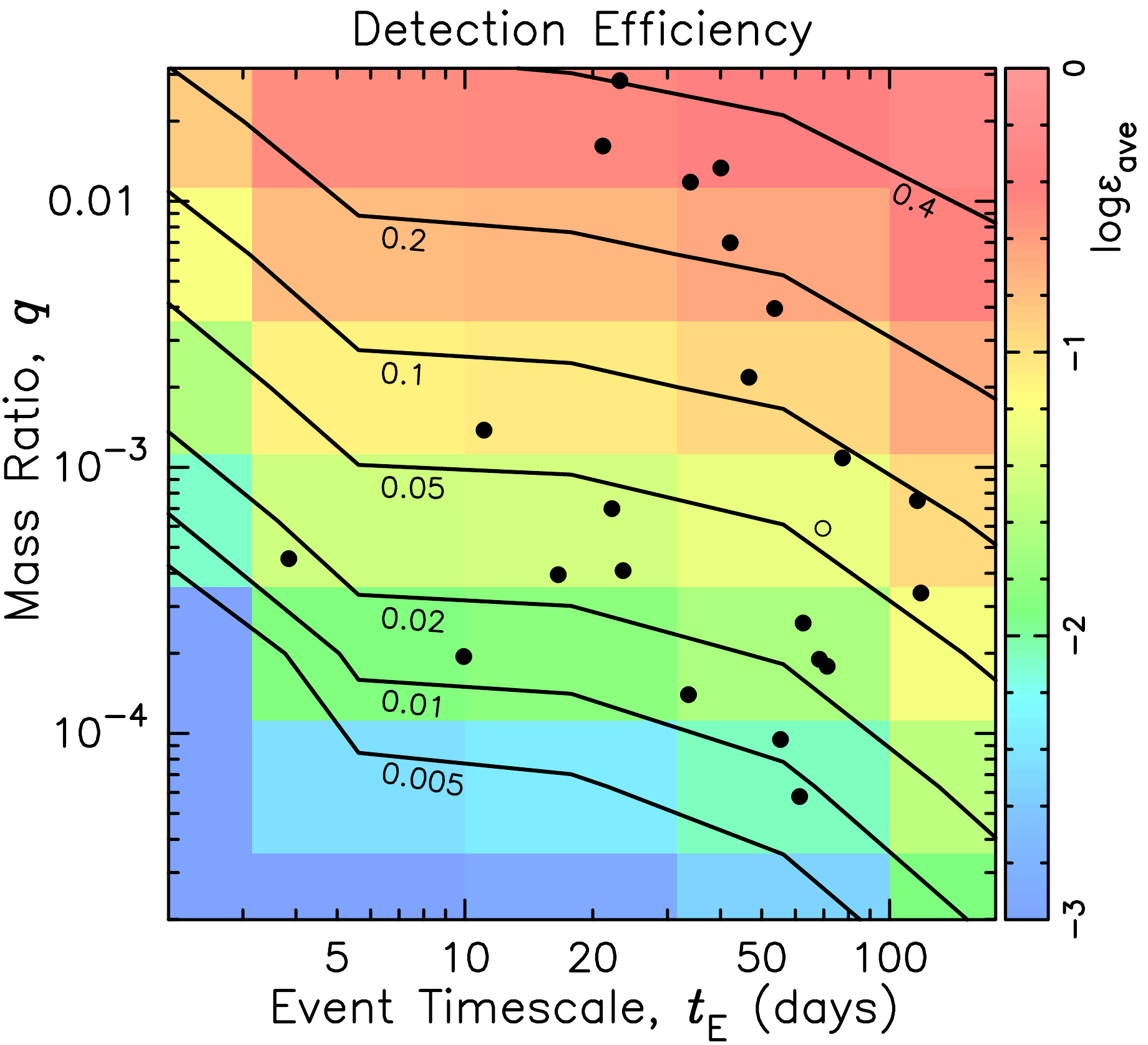}{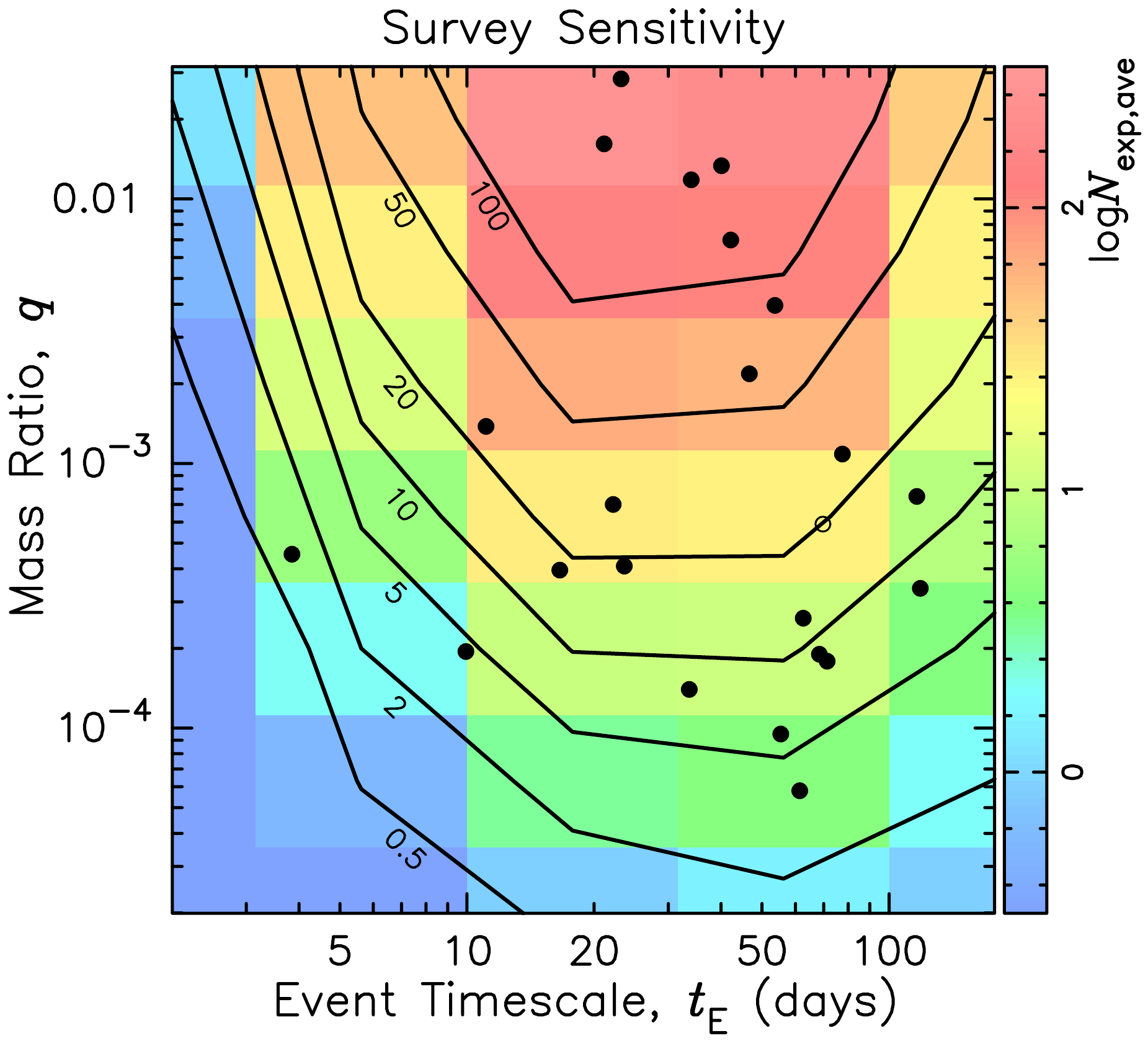}
\caption{
Left: detection efficiencies as a function of mass ratio and $t_{\rm E}$.
For each $0.5\,{\rm dex} \times 0.5\,{\rm dex}$ bin, detection efficiencies are averaged.
The filled circles show the 22 planetary events in this analysis, and the open circle indicates the
ambiguous event. The contours show the averaged detection efficiencies of 40\%, 20\%, 10\%, 5\%, 2\%, 1\% and 
0.5\%, as labeled from top to bottom.
Right: the survey sensitivity, 
$N_{\rm sens}(\log q_{\rm bin}, t_{\rm E}) = \sum_{i=1}^{N_{{\rm bin}}} \epsilon(\log s_i, \log q_i)$,
which is given by the summation of the detection efficiencies for each observed event 
in $0.5\,{\rm dex} \times 0.5\,{\rm dex}$ bins.
While the detection efficiency is highest for very long events, there are few such events, so the
survey sensitivity is highest for events in the $t_{\rm E} \sim 20 - 50\,\rm days$ range.
The contours show the averaged number of expected planets of 100, 50, 20, 10, 5, 2, and 0.5 from top to 
bottom, as the labels indicate.
}
\label{fig:tE_eff}
\end{center}
\end{figure}

In this section, we investigate the planet frequency as a function of $t_{\rm E}$.
The median of $t_{\rm E}$ is $\sim 20$ days for microlensing events toward the Galactic bulge,
but the median of $t_{\rm E}$ for planetary events is $\sim 40$ days, as shown in 
Figure \ref{fig:tE_dist}. This effect is largely explained by the 
higher planet detection efficiency for long $t_{\rm E}$ events, as shown in the left
panel of Figure \ref{fig:tE_eff}. But, it is possible that planets favor long $t_{\rm E}$ events, 
even after correction for the detection efficiency.
If planets favor long $t_{\rm E}$ events, it could imply that cold planets are more common
orbiting massive stars, Galactic disk stars, or both.
Note that it is important that we have selected only evens with robust 
$t_{\rm E}$ measurements in order to avoid systematic errors associated with large
$t_{\rm E}$ uncertainties.

The left panel of Figure \ref{fig:tE_eff} shows the detection efficiencies 
$\log  \varepsilon(\log \,t_{\rm E}, \log \,q)$ in the $t_{\rm E}$ - $q$ plane.
The detection efficiency is averaged over 0.5 dex bins in both $\log \,t_{\rm E}$ and $\log q$ for visualization.
The detection efficiency is larger for long $t_{\rm E}$ and higher $q$ values.
But, the number of events drops for $t_{\rm E} \simgt 60\,$days 
due to the small number of long $t_{\rm E}$ events in our sample.
So our sensitivity to planets in longer events drops, as shown in the right panel of Figure \ref{fig:tE_eff}.
This panel shows the survey sensitivity, which is the sum of the detection efficiency for each
event, in 0.5 dex bins in both $\log \,t_{\rm E}$ and $\log q$. The survey sensitivity is highest
for events with $t_{\rm E} \sim 20 - 50\,\rm days$ where both the detection efficiency and the
number of events are relatively high.

\begin{figure}
\plottwo{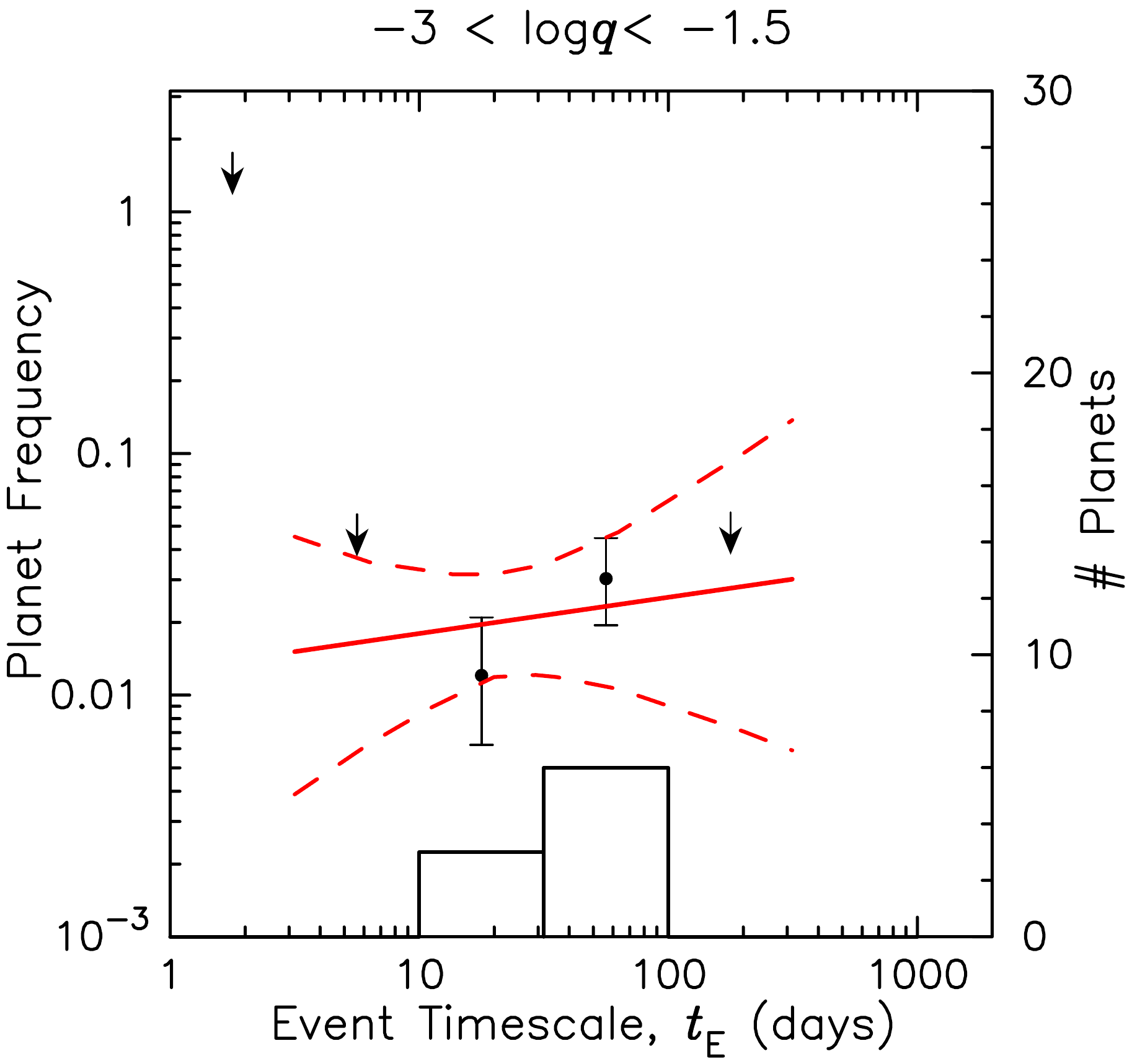}{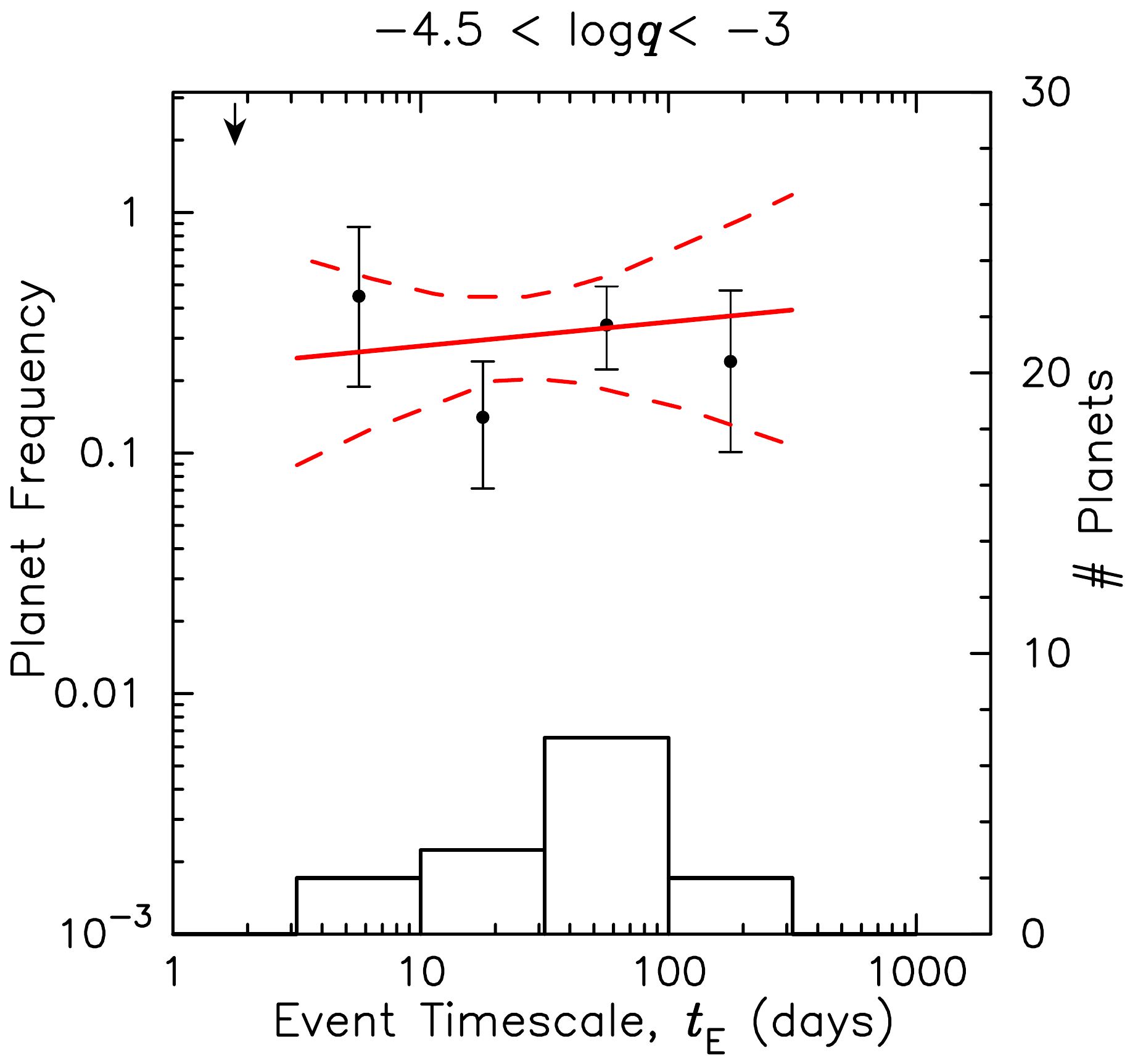}
\caption{
Planet occurrence frequency as a function of $t_{\rm E}$ for high mass
ratio planets ($-3 < \log q < -1.5$) on the left and low-mass ratio planets 
($-4.5 < \log q < -3$)  on the right.
}
\label{fig:tE_func}
\end{figure}

We expect any trend with $t_{\rm E}$ to be relatively weak, because $t_{\rm E}$ depends on
the lens primary mass and distance, as well as the lens-source relative transverse velocity. 
Thus, even a strong trend with host mass of galactocentric distance could be partially 
washed out by the variation of the other variables that $t_{\rm E}$ depends on. 
With only 22 or 23 events, we do not have a great deal of sensitivity to subtle variations
in the planet frequency with $t_{\rm E}$, so we divide the planetary events into two mass 
ratio ranges; the 14 planets with $-4.5 < \log \,q < -3$,  and 9 planets with $-3 < \log \,q < -1.5$.
The planets in the lower mass range have masses less than that of Jupiter mass, and
are mostly cold Neptunes, while the planets in the later mass range are Jupiters and
super-Jupiters. As in Section \ref{subsec:fun_q_sep}, we define $f_{t_{\rm E}}$ as a number of 
planets per star per logarithmic event timescale,
\begin{equation}
f_{t_{\rm E}} \equiv \frac{dN_{\rm pl}}{d\,\log \,t_{\rm E}} = C_{0}\left( \frac{t_{\rm E}}{t_{\rm E\,0}} \right )^{\gamma},
\end{equation}
where $N_{\rm pl}$ is an averaged number of planets around a lens star for the two mass ratio ranges. 
The parameters $C_{0}$ and $\gamma$ are the 
normalization and slope for the assumed power law function, respectively.
We select a pivot point of $t_{\rm E\,0} = 30\,$days. We use a likelihood analysis to 
determine $C_{0}$ and $\gamma$ with an assumption of uniform priors.
The likelihood function is just the poisson probability of finding the observed number 
of events, $N_{\rm obs}$, times the product of the probability of finding events with each 
of the observed event timescale, $t_{{\rm E}\,i}$,
\begin{equation}
\mathcal{L}(C_{0},\gamma) = e^{-N_{\rm exp}}\prod_{i}^{N_{\rm obs}} f_{t_{\rm E}} \varepsilon(t_{{\rm E}\,i});\ \ \ N_{\rm exp} = \int dt_{\rm E} f_{t_{\rm E}} \varepsilon(t_{{\rm E}\,i}), 
\end{equation}
where $\varepsilon(t_{{\rm E}\,i})$ is the detection efficiency in each 0.5 dex $t_{\rm E}$ bin, 
and $N_{\rm exp}$ is the number of events expected for the given $C_{0}$ and $\gamma$ values, 
integrated over $0.5 < \log t_{\rm E} < 2.5$. 
We find that the power law index $\gamma$ is flat for both mass ratio 
ranges: $\gamma = 0.14 \pm 0.39$ for $q > 10^{-3}$, and $\gamma = 0.07 \pm 0.30$ 
for the range with $q < 10^{-3}$, as shown in Figure \ref{fig:tE_func}. So, the 
mass ratio function is flat or slightly increasing with $t_{\rm E}$. The normalizations
are $C_0 = 0.018^{+0.007}_{-0.006}$ for $q > 10^{-3}$ and
$C_0 = 0.28^{+0.09}_{-0.07}$ for $q < 10^{-3}$. A larger survey
will be evidently be needed to measure the $t_{\rm E}$ dependence of the cold
exoplanet mass ratio function.

\section{Discussion}
\label{sec_dis}

\subsection{Comparison with Previous Microlensing Results}
\label{subsec:com_micro}

Previous attempts to measure the exoplanet mass ratio or mass function using microlensing
data have used more limited samples than we have used in this paper. The first attempt
was the measurement by \citet{sum10} of the slope of the exoplanet mass ratio function using
only relative and not absolute detection efficiencies. \citet{sum10} used the 10 
microlensing discoveries known at that time to estimate logarithmic mass ratio function slope of
$n = -0.7\pm 0.2$, which is consistent with our value of $n = -0.90^{+0.15}_{-0.17}$
for the mass ratios $> q_{\rm br}$. 
But, without the break in the mass ratio function, it appears that the events with $q < q_{\rm br}$
have flattened their mass function slope.

\begin{figure}
\begin{center}
\plotone{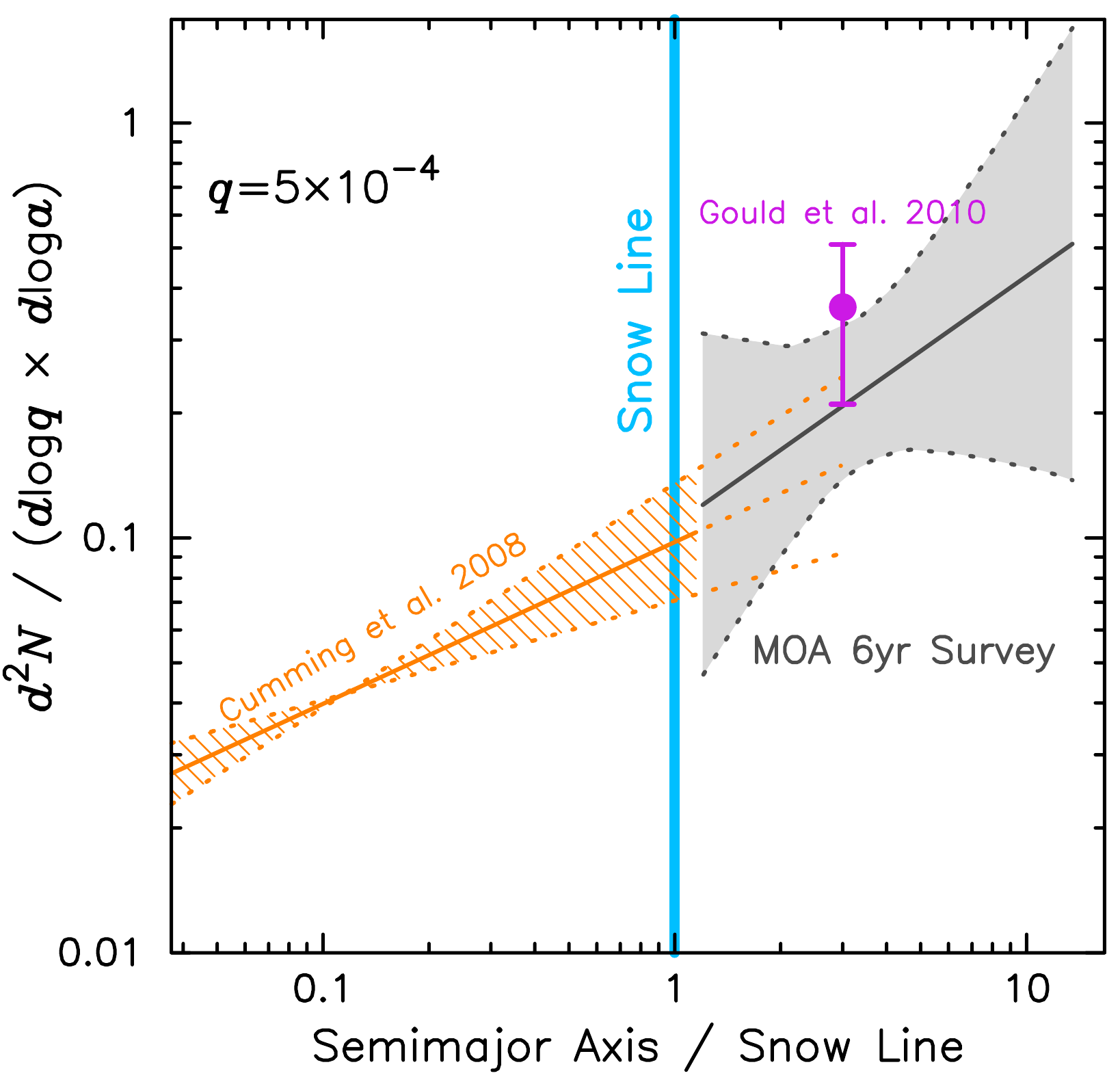}
\caption{
Planet frequencies at the mass ratio of $q=5\times10^{-4}$ as a function of semi-major axis, which is normalized by the snow line $\sim 2.7(M/M_\odot) {\rm AU}$.
Microlensing results are compared with the radial velocity result of \citet{cum08}. 
We assume that the exoplanet frequency scales with the mass ratio, $q$, and the
separation compared to the snow line. The \citet{gou10} and our pivot point ($s=1$) is plotted at 
three times the snow line.
}
\label{fig:a_func}
\end{center}
\end{figure}

\citet{gou10} used 13 high magnification events containing 6 planetary signals for the first
microlensing analysis with absolute detection efficiencies. These events were selected
based on their observation by the $\mu$FUN microlensing follow-up collaboration
They did not solve for a slope
with mass ratio, and they find $d^2N_{\rm pl}/(d\log q\ d\log s) = 0.36\pm 0.15$
at an average mass ratio of $q = 5\times 10^{-4}$, 
assuming a distribution that is flat in both $\log q$ and $\log s$. As shown in Figure~\ref{fig:a_func},
the $\mu$FUN result is a factor of $\sim 2.0$ higher than our result when evaluated
at the central $q$ for the $\mu$FUN analysis. Our value is 1.2$\,\sigma_{\mu\rm FUN}$ 
below the $\mu$FUN result, and less than 1-$\,\sigma$ when the combined MOA and
$\mu$FUN error bars are used. However, our mass ratio function has a different functional
form than theirs. So, a better comparison would be to integrate our mass ratio functions
over a range of mass ratios centered at $q = 5\times 10^{-4}$. In order to avoid the
complication associated with the break, we select the range extending from $q_1 = 2\times 10^{-4}$
to $q_2 = 1.25\times 10^{-3}$, a factor of 2.5 in each direction from the central value.
Our mass ratio function average over this $q_1$-$q_2$ range is 0.226, which compares
to the $\mu$FUN value of $0.36\pm 0.15$. So, our value is 37\% lower.

Our mass ratio function is fully consistent with the earlier $\mu$FUN measurement, so there is
no need to identify a bias to explain the fact that our value is lower. 
However, we have identified such a bias, nonetheless. This is a ``publication date" bias. That is,
the publication date for the $\mu$FUN sample was not decided in advance. Instead, it was selected,
in part, because the authors judged that they had discovered enough planets for such a paper. Such
a decision is more likely after a few years or two with a higher than average planet detection rate. 
However, we have identified a possible bias, nonetheless.
\citet{gou10} found 6 planets in 4 years of data for a rate of 1.5 per year, 
but the following 6 years of observations yielded
only 3 planets with $\mu$FUN data \citep{bac12pl,yee12,fuk15} that pass the criteria of \citet{gou10},
when we would have expected 9.
Event OGLE-2013-BLG-0341 does not qualify for this sample because of its stellar
binary signal is much stronger than the signal of the planet \citep{gou14}. There are also several recent strong
planetary events with important $\mu$FUN data that are excluded from this list of planets because they
have $|u_0| > 0.005$ \citep{miy11,han13,yee14}.
So, it is likely that an update of the \citet{gou10} analysis will yield a result closer to our MOA result.
The possibility of a similar bias affecting our own analysis seems to be smaller because our
sample is larger.

While the previous analyses of \citet{sum10} and \citet{gou10} dealt with exoplanet mass ratio, $q$,
as our analysis does, \citet{cas12} attempted to derive the exoplanet mass function. They find 
$d^{2}N_{\rm pl}/(d\,\log a\ d\,\log M) = 0.24_{-0.10}^{+0.16}(M/M_{\rm Sat})^{-0.73\pm0.17}$, 
where $a$ is semi major axis, $M$ is the planetary mass, and $M_{\rm Sat}$ is the mass of 
Saturn, $M_{\rm Sat} = 95.2 M_{\oplus}$. The PLANET Collaboration \citep{cas12} analysis 
includes constraints from 
\citet{sum10} and \citet{gou10}, and it uses a Bayesian analysis, rather than mass and 
distance measurements to convert from mass ratios to masses. This can be somewhat
misleading, as it makes the assumption that exoplanet properties do not depend on 
the mass or distance of the host star. Since no mass measurements are used, we feel that
it is more sensible to convert the \citet{cas12} result back to mass ratios instead of masses.
This can be done trivially because a power law mass function transforms into a power
law mass ratio function. (This applies for every mass, so it applies for all masses.)
However, we must note that \citet{cas12} do not correctly estimate the median mass that
their survey probes for planets. They estimate the median mass corresponding the event
with the median $t_{\rm E}$ in the detection efficiency sample. However, the planet detection
efficiency is higher for longer $t_{\rm E}$ events, so the sensitivity is increased for longer events.
Since the planet occurrence rate is only weakly sensitive on $t_{\rm E}$, as shown in 
Figure~\ref{fig:tE_func}, we can use the median value of $t_{\rm E}$ for the detected planets,
which is $t_{\rm E} = 42.5$ days for the 23 events in the MOA sample, as well as the 
combined MOA+$\mu$FUN+PLANET sample discussed below. With the Galactic model
of \citep{ben02}, this corresponds to a median mass of $0.59M_\odot$. This is more
appropriate median mass for the PLANET sample, so we use it to convert the
\citet{cas12} mass function back into a mass ratio function.
This converted result (which is used for only this comparison, not for the following subsection) is compared to our result and the $\mu$FUN analysis 
in Figure~\ref{fig:compRV}. The PLANET result is obviously consistent with our
mass ratio function.

The most recent statistical analysis of microlensing exoplanet data is the \citet{shv15}
analysis of the combined MOA, OGLE and Wise 2011-2014 microlensing survey
data set. This analysis is rigorous in the sense that the detection efficiency for 
binary and planetary microlensing events is calculated, 
but the analysis of the detected binary and planetary microlensing events is incomplete.
Some of the binary and planetary events have had a complete analysis done in other papers,
and the results of these published analyses are used when they are available. 
But, for the other binary and planetary events, the light curve analysis has been
limited to a coarse grid in $q$ and $s$. This results in light curve models that obviously, by eye, do not
fit the data for many of their binary and planetary events, including OGLE-2011-BLG-0481, 
OGLE-2011-BLG-0974, OGLE-2012-BLG-0442, OGLE-2014-BLG-0257,
OGLE-2014-BLG-0572, OGLE-2014-BLG-0676, OGLE-2014-BLG-1255, and
OGLE-2014-BLG-1720, which could lead to incorrect statistical inferences about planetary system properties.
For example, OGLE-2011-BLG-0481 is a stellar binary, but the model presented
in \citet{shv15} is a planetary model with $q\sim 0.014$. OGLE-2014-BLG-0676 is correctly
classified as a planetary event, but \citet{shv15} estimate a mass ratio of 
$q \sim 1.4\times 10^{-3}$ whereas the full analysis by \citep{rat16} finds
$q = 4.78 \pm 0.13 \times 10^{-3}$, a difference of a factor of 3.4.
They used 9 planets including the above one to find a mass ratio slope of $dN/d\,\log q = q^{-0.50 \pm 0.17}$, which is
only marginally consistent with our slope of $dN/d\,\log q = q^n$ with
$n = -0.91^{+0.15}_{-0.17}$.
They also find that the average number of planets per star is $0.55^{+0.34}_{-0.22}$
with $10^{-4.9} < q < 10^{-1.4}$ and $0.5 < s < 2$. This compares to 0.43 planets per
star with the same range of $q$ and $s$ values for our combined mass ratio
function, discussed in the next section.

\subsection{Combined Microlensing Mass Ratio Function Analysis}
\label{subsec:comb_ML}

Our analysis includes 22 planetary events and one likely planetary event, which compares
to 6 and 8 events used in the previous statistical analyses of exoplanet microlensing 
data sets \citep{gou10,cas12}. However, 7 of the 8 events used in these previous
analyses are not included in our sample. (The planetary event, OGLE-2007-BLG-349
is included in all three samples.)
Thus, we can increase our sample to 29-30 
planets by adding these samples to the MOA sample. This is straightforward to do
because \citet{gou10} and \citet{cas12} have made their  
survey sensitivity functions, $S(\log s, \log q)$ and detection efficiencies,
$\epsilon(\log s, \log q)$, available to us.
(To compute the survey sensitivity as a function of the semi-major axis and planet mass, \citet{cas12} converted 
their detection efficiencies as a function of $\log s$ and $\log q$ to those of physical parameters. 
We use their detection efficiencies as a function of $\log s$ and $\log q$ (A. Cssan, private communication).)
We then simply multiply the likelihood functions, Equation~(\ref{eq:q-s}), for the three different
surveys together to get our final likelihood function. The only complication is the 
overlap between the three surveys. We deal with this by removing the 2007 season
from the PLANET sample and removing the events in the $\mu$FUN analysis
from the MOA sample. We make these choices because the MOA 2007 season
has a higher sensitivity than the PLANET 2007 season and because the inclusion
of the follow-up data gives the $\mu$FUN analysis higher sensitivity for the 
13 events in their sample, despite their higher detection threshold.

\begin{figure}
\begin{center}
\plotone{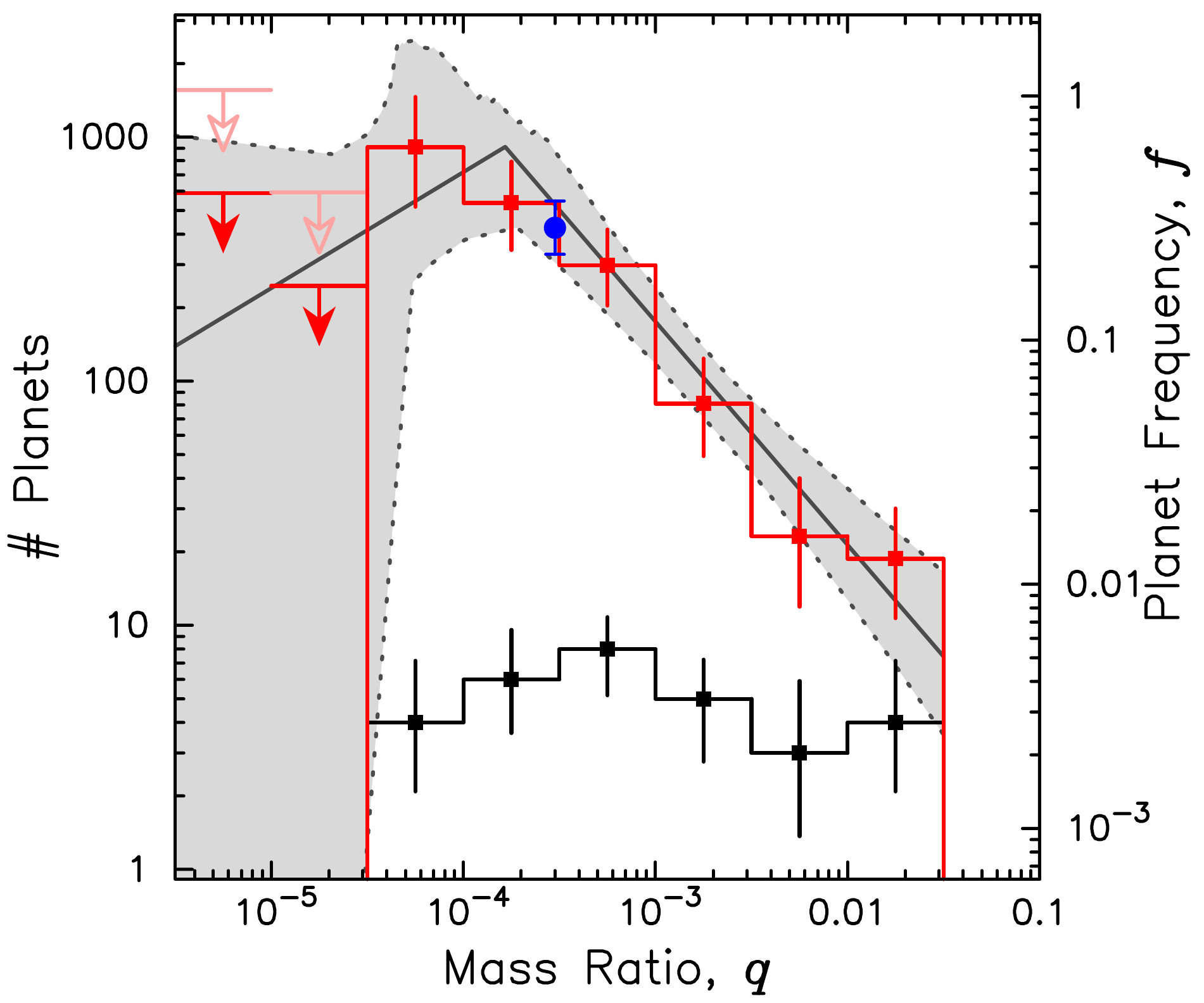}
\caption{
Mass-ratio distribution, as in Figure~\ref{fig:q_func}, for the combined microlensing data set.
The black histogram is the number of detected planets per bin, and the red histogram is corrected
for detection efficiencies.
}
\label{fig:q_func_all}
\end{center}
\end{figure}

\begin{figure}
\begin{center}
\plotone{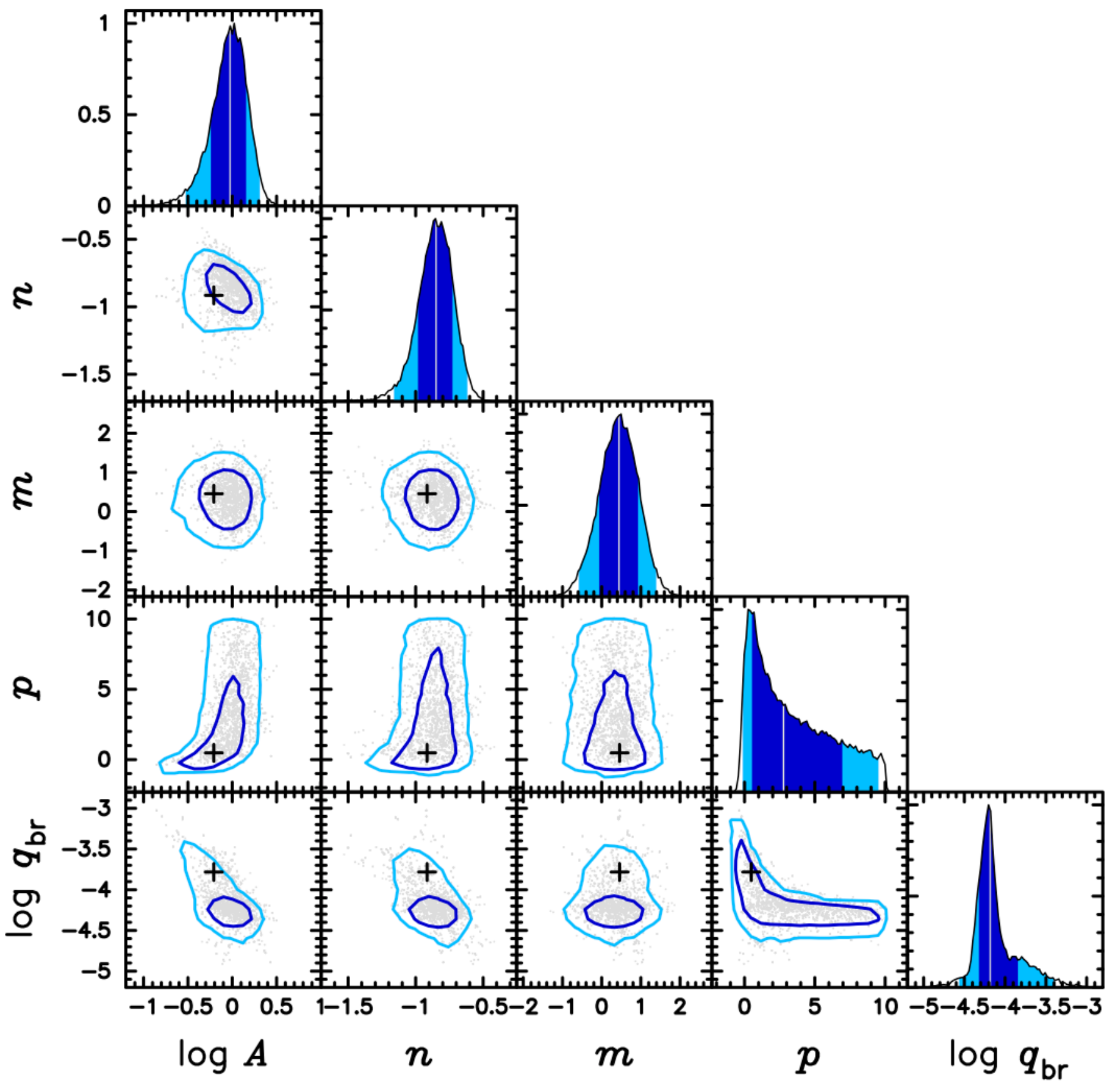}
\caption{
Likelihood contours for the broken double power law function, $f =  A \left[\left(\frac{q}{q_{\rm br}} \right)^{n}\Theta(q-q_{\rm br}) +  \left(\frac{q}{q_{\rm br}} \right)^{p}\Theta(q_{\rm br}-q)\right] \left(\frac{s}{s_{0}}\right)^{m}$
for the full MOA+$\mu$FUN+PLANET sample.
The contours indicate the 68\% and 95\% confidence intervals, and the crosses indicate the
best fit parameters (which are also listed in Table~\ref{tab:expl_bmrf}).
}
\label{fig:mrf_contALL}
\end{center}
\end{figure}

The results of this combined analysis are summarized in Tables~\ref{tab:expl_bmrf}
and \ref{tab:expl_mrf}, as well as Figures~\ref{fig:q_func_all} and \ref{fig:mrf_contALL}.
The mass ratio function parameters from this combined sample are very
similar to the parameters from the MOA-only sample. Again, the uncertainties in mass ratio and separation 
for the additional planets from \citet{gou10} and \citet{cas12} are negligible. They change the best fit model 
by 10 \% of the uncertainties for each parameter. With this combined sample, 
the best fit broken power-law model is a better fit than the best single power law
model by $\Delta \chi^{2} = 12.2$ (with 2 additional parameters). This implies a $p$-value
of 0.0022, and a Bayes factor of 27, implying that the broken power-law model is favored
over the single power-law model by a factor of 27.
The normalization is increased by $5\,$\% or $\sim 0.13\,\sigma$ when $q_{\rm br}$ is fixed,
and the $q_{\rm br}$ value is decreased somewhat when $q_{\rm br}$ is set as a free parameter.
As expected, the larger sample decreases the error bars slightly.

We can also use this combined sample to look at the number of planets implied by 
our mass ratio function. Microlensing has been traditionally thought to be sensitive
to planets in a so-called ``lensing zone" \citep{gou92,ben96}, which roughly spans
the range $0.5 < s < 2$. If we integrate our best fit mass ratio function for the
combined sample (shown in Figures~\ref{fig:mrf_contALL} and Table~\ref{tab:expl_bmrf}), 
we find 0.48 planets per stars inside the ``lensing zone\rlap." However, some of these
planets have $q < 5\times 10^{-5}$, which is below the lowest mass ratio for any planet
in our combined sample. If we consider only planets in the range of our known
microlens planets, with $q >5\times 10^{-5}$, we find 0.33 planets per star. On the 
other hand, 10\% of the combined microlens exoplanet sample are located outside
the nominal $0.5 < s < 2$ ``lensing zone\rlap." To include all the planets in our sample,
we must use an extended ``lensing zone" spanning the range $0.3 < s < 5$. Our
mass ratio function predicts 0.79 cold planets per star with $q >5\times 10^{-5}$ and
1.26 cold planets per star if we extrapolate our mass function down to $q = 0$.

\subsection{Comparison with Radial Velocity Results}
\label{subsec:com_rv}

Our microlensing mass ratio function results are more readily compared with
the results of RV surveys than transit surveys because both 
microlensing and the RV method detect exoplanet masses rather
than the radius, which is detected by the transit method. We consider 
five different RV studies in refereed journals that report statistical
results \citep{cum08,may09,how10,joh10occ,how10,bon13,mon14}. All of
these studies have smaller exoplanet sample than our combined sample
except for \citet{cum08} and possibly \citet{may09}, and none of them
have an exoplanet sample more than a factor of two larger than our sample.

\begin{figure}
\begin{center}
\plotone{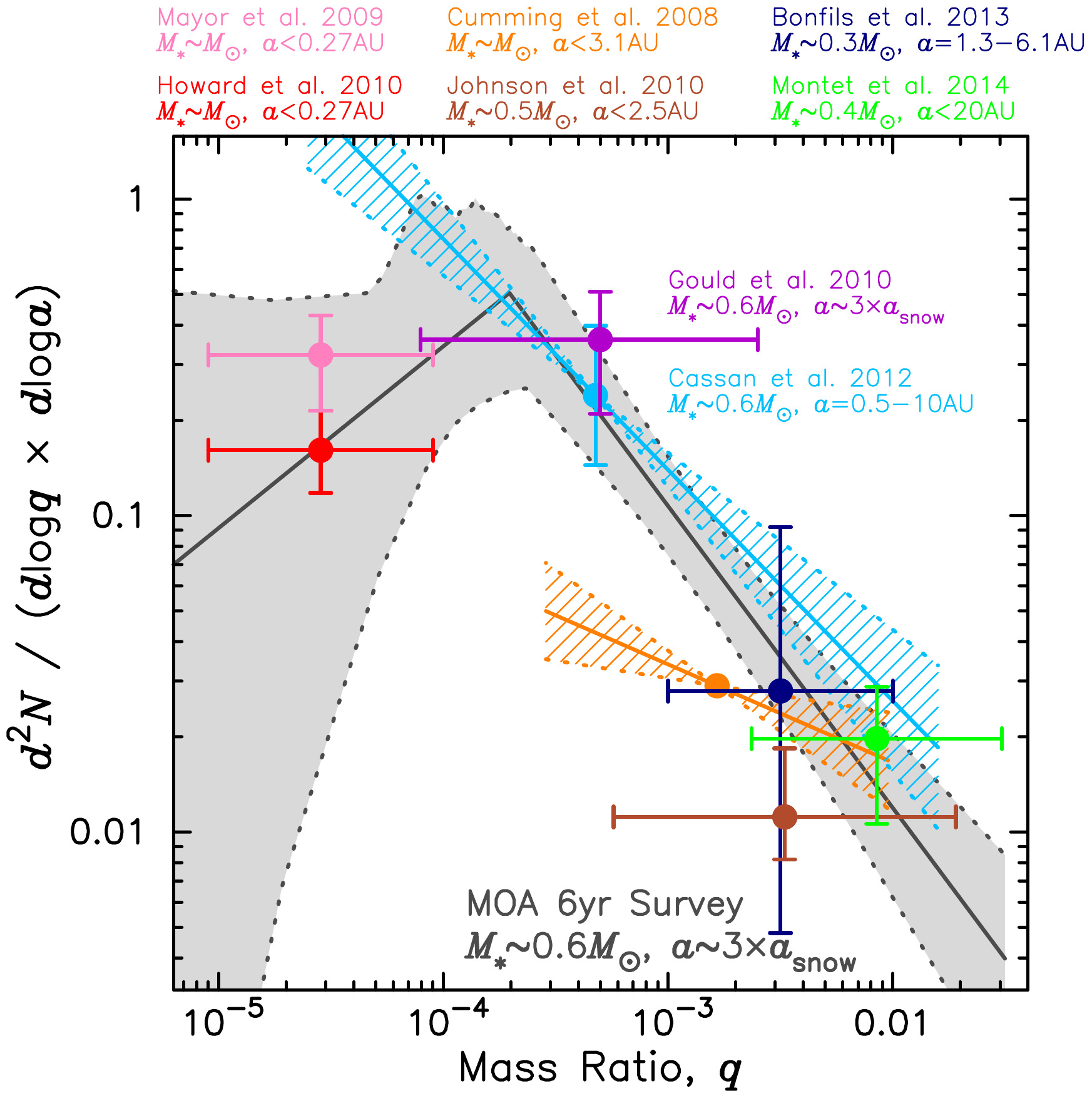}
\caption{MOA exoplanet frequency result from Figure~\ref{fig:q_func}
as a function of mass ratio is compared with previous microlensing and RV results.
The reference, type of primary star, and semi-major axis are labeled in the same color as 
the points, error bars and shaded regions in the main body of the figure.
}
\label{fig:compRV}
\end{center}
\end{figure}

\begin{figure}
\begin{center}
\plotone{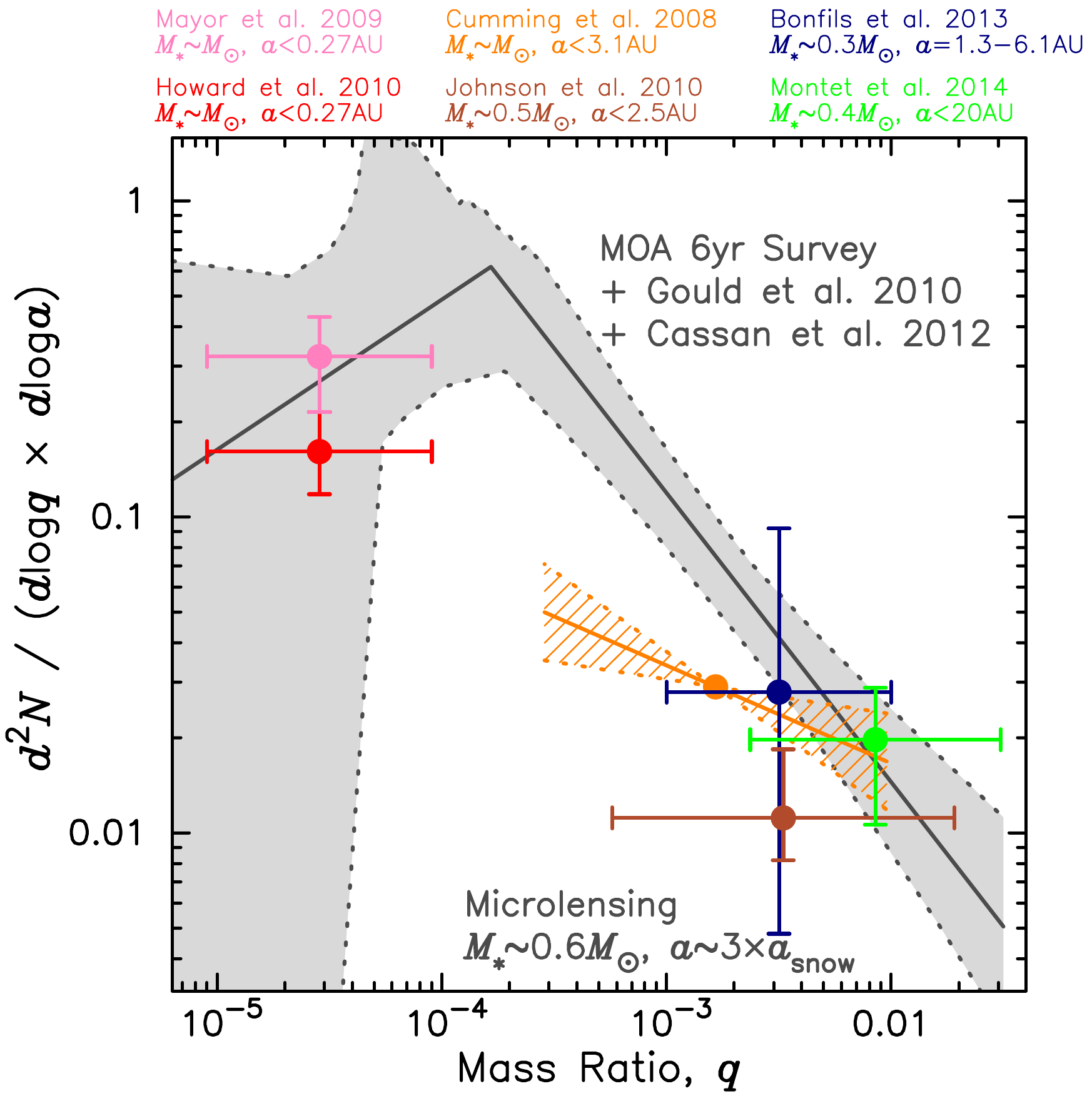}
\caption{
Combined MOA+PLANET+$\mu$FUN exoplanet frequency result from Figure~\ref{fig:q_func_all}
as a function of mass ratio is compared with previous microlensing and RV results.
The reference, type of primary star, and semi-major axis are labeled in the same color as 
the points, error bars and shaded regions in the main body of the figure.
}
\label{fig:compRV_all}
\end{center}
\end{figure}

Two of these studies \citep{may09,how10} focus on planets orbiting stars
of $\sim 1\,M_\odot$ in short period orbits (period $< 50\,$days), and so they
have no overlap with our planetary microlens sample. However their
sensitivity does extend down to $q \sim {\rm a\ few} \times 10^{-5}$,
where these results are consistent with the mass ratio function measured
with our microlensing at much larger orbital separations, as shown in 
Figures~\ref{fig:compRV} and \ref{fig:compRV_all}. 

\begin{figure}
\begin{center}
\plotone{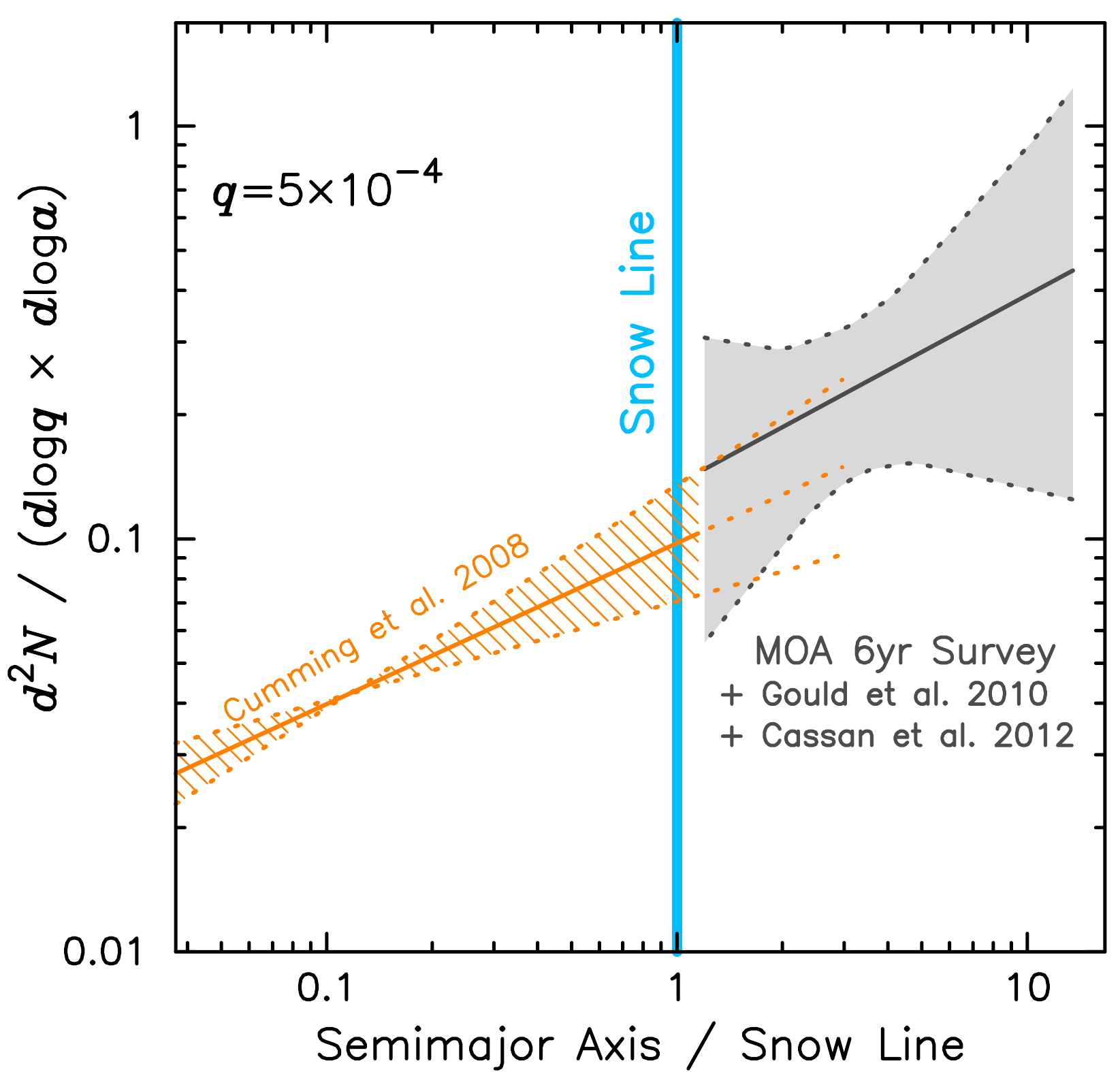}
\caption{
Same as Figure \ref{fig:a_func}, but the result of the full microlensing sample is compared to \citet{cum08}.
separation compared to the snow line. 
}
\label{fig:a_func_all}
\end{center}
\end{figure}

The study by \citet{cum08} had the largest sample (48 planets) and covered periods 
of $P < 2000$ days (corresponding to $a \simlt 3.1 {\rm AU}$ for solar mass host stars),
but they have little sensitivity to planets less massive than Jupiter in wide orbits.
So, they were sensitive primarily to planets with shorter period orbits than the 
planets probed by the microlensing method, and their host stars are more
massive than most of the planets found by microlensing. Nevertheless,
our results do agree well with their slope with separation, as indicated in
Figures~\ref{fig:a_func} and \ref{fig:a_func_all}. While the \citet{cum08}
slope with semi-major axis was consistent with the microlensing result
of \citet{gou10}, the agreement with our result is somewhat better. However,
our constraint on the slope of the mass ratio function is rather weak,
$m \simeq 0.5\pm 0.5$ for the complete sample, so it is consistent with a
logarithmically flat distribution ($\ddot{\rm O}$pik's law) at 1-$\sigma$.
This is due to the relatively
narrow range of planetary separations that most microlensing observations are
sensitive to, as well as the close-wide degeneracy that occurs for high magnification
events that are sensitive to planets at a wide range of separations.

In contrast, microlensing is sensitive to a large range of exoplanet mass ratios.
Our detected planets span a range of 500 in mass ratio, and we have significant
sensitivities at masses below the planet with the lowest observed mass ratio.
The \citet{cum08} result does not match our more precise measurement
of the dependence of the mass ratio function on $q$ at $q \simlt 10^{-3}$, where their
value is lower than our measurement, as our mass ratio function rises more 
steeply towards lower mass ratios than theirs does, as indicated in
Figures~\ref{fig:compRV} and \ref{fig:compRV_all}.
This may be an indication that there are more cold Neptunes beyond the
snow line that have undergone little or no migration than there are at 
$\sim 1\,$AU. 

The RV studies of \citet{joh10occ}, \citet{ bon13} and \citet{mon14} 
focused on M-dwarf primaries, although with rather small samples of 5, 11, and 14 planets,
respectively. \citet{joh10occ} finds a significantly lower occurrence frequency than
we find, but this is largely due to the limited mass and separation sensitivity of
their RV survey. In contrast, using some of the same data, \citet{mon14} 
find an occurrence rate consistent with our measurements. (See Figures~\ref{fig:compRV} 
and \ref{fig:compRV_all}.) They have been
much more careful to consider the RV selection effects and the possible
signals of planets with periods longer than the survey. 
\citet{bon13} also searched for the planets orbiting M-dwarfs using the HARPS instrument, and 
found planets spanning a range of 300 in mass and 150 in semi-major axis, but only two of
these planets are in the range of periods, $10^{3}$ -- $10^{4}$ days, where microlensing is
sensitive. However, only one of the two planets were confirmed by the HARPS survey (as pointed out in \cite{cla14b}). So, we consider only one planet discovery to plot the \citet{bon13} results in 
Figures~\ref{fig:compRV} and \ref{fig:compRV_all}.

A recent study by \citet{cla14a,cla14b}, simulated 
RV observations of the planet population found by microlensing showed that 
the number of RV detections expected by the simulation agrees with the actual RV 
observation of \cite{bon13}. However, this was a comparison to the \citet{cas12}
mass function, so it is not a direct comparison with our results.

High contrast imaging is a promising
method to determine the frequency of exoplanets in wide orbits, but their sensitivity
is currently limited to self-luminous planets, and their published statistical results
tend to be upper limits \citep{ram13}. So, this method does not yet have
a great deal of overlap with the primary sensitivity region for the microlensing
method at 2-$3\times$ the snow line.

\subsection{Comparison with {\it Kepler} Results}
\label{subsec:com_kep}

By far, the largest statistical samples of planets with well-characterized detection efficiencies
come from the {\it Kepler} mission \citep{pet13a,for14,dre15}, but {\it Kepler} provides only a direct
measurement of the planet radius and not mass for the vast majority of planets discovered.
Not only that, but the {\it Kepler} planets for which both the radius and mass are measured are
biased on sample of planets. Masses are more readily determined by RVs for more 
massive planets in short period orbits, but the smaller planets are considered to be more
interesting. So, much observing time is devoted to planets that are thought to have barely
detectable RV signals. Transiting planet masses can also be determined
by transit timing variations, but many of the transit timing variation measurements 
are made for an unusual class of low-density
planets that are tightly packed in highly co-planar orbits \citep{lis13,jon14}. 
These planets might have different compositions and therefore a different
mass-radius distribution from planets in more sparsely packed, less co-planar systems.
Also, attempts to determine the mass-radius relation have generally not spanned the
full range of planet masses. At masses near a Jupiter-mass and above, the physics
of electron degeneracy pressure conspires to make the planet radius nearly independent
of the planet mass (but still dependent on the planet composition). 

\begin{figure}
\begin{center}
\plotone{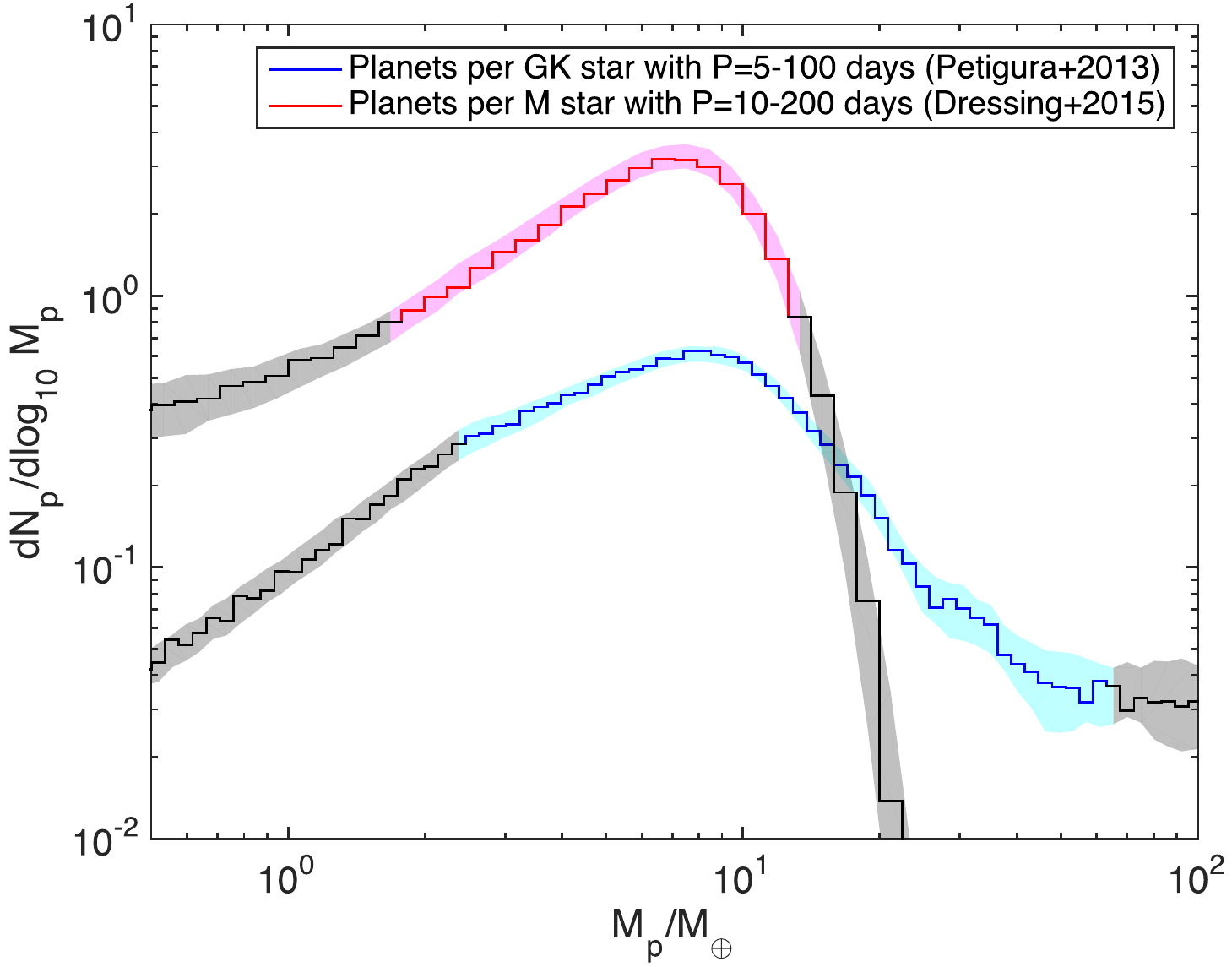}
\caption{
Planet frequencies for M- and GK-dwarf hosts as a function of mass from the {\it Kepler} mission,
based upon the \citet{dre15} and  \citet{pet13a,pet13b}  exoplanet radius functions
for M and GK stars, respectively. These have been converted to masses using the
probabilistic mass-radius relation of \citet{wol15}. The red and blue colored curves indicate
the reliable region of these curves, and the gray colored regions indicate areas where the 
input radius functions or the mass-radius relation are thought (by the respective
authors) to be unreliable.
}
\label{fig:exmp_kepGKM}
\end{center}
\end{figure}

As a result of these considerations, we have limited our comparison to {\it Kepler} to 
a simple comparison of the masses of the mass function breaks or peaks seen
in each data set. For {\it Kepler}, we use the exoplanet radius functions of \citet{pet13a}
for G and K host stars and \citet{dre15} for M-dwarf hosts. To convert to masses,
we use the probabilistic mass-radius relation of \citet{wol15} in both cases.
The exoplanet mass functions derived from \citet{pet13a} and
 \citet{dre15} are shown in Figure~\ref{fig:exmp_kepGKM}. Only the red and blue
 colored part of the curves in this figure are reliable.
Both of these exoplanet mass functions 
have a similar shape in the region where both are reliable, $2.5M_\oplus < M_{\rm pl} < 13M_\oplus$.
The mass functions for M and G, K host stars have apparent peaks 
at $\sim 6 M_\oplus$ and $\sim 8 M_\oplus$, respectively.

Because there are few microlens planets below the mass ratio function break at $q_{\rm br} \sim 10^{-4}$,
we do not have a very precise measure of $q_{\rm br}$. From Table~\ref{tab:expl_mrf},
we see that the 1-$\sigma$ range is roughly 0.5-$2\times 10^{-4}$. Since the typical host
star mass is $\sim 0.6M_\odot$, for our detected planets (in both the MOA and Combined samples) 
$q_{\rm br} = 10^{-4}$ corresponds to $\sim 20M_\oplus$,
which is just above the mass of Neptune. 
Thus, the break in the mass function beyond the snow line is likely to occur at
a mass in the range 10-$40M_\oplus$. This is likely to be somewhat higher
than the apparent mass function peaks in shorter period orbits, as estimated
from {\it Kepler} data, but we will need a larger sample of low-mass microlens
planets to confirm that the break in the mass function at higher masses
beyond the snow line.

We can also compare the absolute numbers of planets found by {\it Kepler} at separations
well inside the snow line to our values well outside the snow line. We focus on the 
planets near the peak of the mass ratio function and the {\it Kepler} radius functions.
From the \citet{pet13a} sample, we consider planets in the period range
$25\,{\rm days} < P < 200\,{\rm days}$. This spans a factor of 8 in period, which is
equivalent to a factor of 4 in separation, by Kepler's 3rd law. Using the mass-radius
relation we select lower limits for $R_p > 2 R_\oplus$ for the  \citet{pet13a} sample 
as equivalent to a lower limit of $q > 2.6\times 10^{-5}$ in our analysis. We find
0.27 planets per star when we sum the  \citet{pet13a} bins between
$25\,{\rm days} < P < 200\,{\rm days}$ with $R_p > 2 R_\oplus$. The 
equivalent range for our exoplanet mass ratio function (Equation~(\ref{eq:expl_mrf})
with the parameters from the $q_{\rm br}$ free-All
column from Table~\ref{tab:expl_bmrf}) is $2.6\times 10^{-5} < q < 0.03$ and
$0.5 < s < 2$. Integrating over this range yields 0.38 planets per star, suggesting
that the peak of the exoplanet mass function of cold planets might have $\sim 41\,$\%
more planets than an equivalent size region near at the peak of the warm planet
distribution for G and K stars. We can also do a similar calculation with the tabulated radius 
function of \citet{dre15} for early M-dwarf host stars. To match the bins of their
tabulated radius distribution, we select $0.45 < s < 2.22$ and $2.9\times 10^{-5} < q < 0.03$
to span the same logarithmic range in separation and mass ratio as 
planets at the peak of the M-dwarf exoplanet radius function with
$10\,{\rm days} < P < 110\,{\rm days}$ with $R_p > 2 R_\oplus$.
Integrating over this $(q,s)$ gives 0.43 planets per star found by
microlensing, as compared to a whopping 3.03 planets per star near
the peak of the early M-dwarf radius function. This dramatic difference between
the height of the peak in the radius function is also seen by \citet{bur15}, but
they also point out that some potentially large systematic errors may
still exist in these analyses.

\subsection{Masses of Host Stars}
\label{subsec:masses}

The masses of the lens stars are not known for the vast majority of microlensing
events in our sample, but the situations is somewhat better for the lens stars
found to host planets. For most planetary events, finite source effects are seen, and
this allows the angular Einstein radius to be determined. When the angular Einstein
radius is known, the lens mass can be determined in those cases where the microlensing
parallax effect can be measured. This has been done for 6 events in our sample,
MOA-2007-BLG-192 \citep{ben08}, OGLE-2007-BLG-349/MOA-2007-BLG-379 \citep{ben16cir}, 
MOA-2009-BLG-266 \citep{mur11}, MOA-2010-BLG-117 \citep{ben16bin},
MOA-2010-BLG-328 \citep{fur13}, and MOA-2011-BLG-197/OGLE-2011-BLG-0265
\citep{sko14}, as well as one event with two planets, OGLE-2006-BLG-109 \citep{gau08,bennett-ogle109},
from the \citet{gou10} sample. Alternatively, it is also possible to determine lens
mass with a measurement of the angular Einstein radius or microlensing parallax
along with the measurement
of the host star brightness. With the help of an empirical or theoretical mass-luminosity
relation, the host star brightness can be turned into a lens mass measurement.  This has
been done using microlensing parallax for OGLE-2012-BLG-0950/MOA-2012-BLG-527
in our sample \citep{kos16}. It also has been done using the angular Einstein radius for 
OGLE-2005-BLG-169 \citep{ben15hst,bat15} and OGLE-2006-BLG-109 
\citep{gau08,bennett-ogle109}
in the \citet{gou10} sample and for OGLE-2005-BLG-071 \citep{don09ob05071} in the \citet{cas12} sample.
The observations that detect the lens are often taken many years after the event, as this
allows us to check that the putative lens star motion is actually consistent with being the lens.
We plan to include these mass constraints in a future analysis.


The host star masses in the planetary events in our sample mainly range from G-dwarfs to late M-dwarfs.
The primary lens masses in MOA-2011-BLG-262 \citep{ben14} could be a Jupiter mass object, although a low
mass star is more likely.
Also, the primary lenses for OGLE-2007-BLG-224/MOA-2007-BLG-163 \citep{gou09} and
MOA-2011-BLG-274 \citep{fre15}, both of which are single lens events and used to estimate the detection 
efficiency, are very likely to be brown dwarfs or possibly a planetary mass
object in the case of MOA-2011-BLG-274.
Thus, conservatively, the host star mass in our analysis is ranging over late-type stars extending to 
both G-dwarfs and brown dwarfs, or even planetary mass objects.


\section{Summary}
\label{sec_sum}

We have reported the planet frequency as a function of planet-to-star mass ratio, 
separation, and event time scale, derived from the microlensing survey data by 
MOA-II telescope in 2007-2012. This sample included 1474 well sampled microlensing
light curves, and we find signals of 22 planets and one likely, but not certain, planet
(see Section~\ref{sec_eve_sel}). Our sample probes the planetary systems of
stars that can serve as lenses for gravitational microlensing surveys towards the
Galactic bulge, so it is an average over stars that orbit inside the Solar Circle.
It is possible that this population is somewhat different from the neighbors that can be
probed by the other detection methods.

Our results are consistent with the previous statistical microlensing analyses of
\citet{sum10}, \citet{gou10} and \citet{cas12}, but we
find that, contrary to previous analyses, the data are not consistent with a single power-law
mass function in the mass ratio, $q$. Instead, we find a strong change in the slope of the
mass ratio function at $q_{\rm br} \sim 10^{-4}$, which we model as a broken power-law mass function,
given by Equation~(\ref{eq:expl_mrf}) or
\begin{eqnarray}
\frac{d^{2} N_{\rm pl}}{d\,\log q\ d\,\log s} &=& A \left(\frac{q}{q_{\rm br}} \right)^{n} s^{m}\ \ \ {\rm for}\ q > q_{\rm br}  \nonumber \\
&=& A \left(\frac{q}{q_{\rm br}} \right)^{p} s^{m}\ \ \ {\rm for}\ q < q_{\rm br},
\end{eqnarray}
with the best fit parameters given in Table~\ref{tab:expl_bmrf} and the Markov chain medians
and error bars given in Table~\ref{tab:expl_mrf}. We present results for the 23-planet
MOA sample, and the full sample of 30 planets (including one somewhat ambiguous event).
The $p$-values for the detection of the break in the mass ratio function are 0.0030 and
0.0022 for the MOA and full samples, respectively. A more conservative estimate of the need for
the power-law break in the mass function ratio is given by the Bayes factors of 21 and 27 favoring
the broken power law for the MOA and full samples. These Bayes factors imply that the
broken power-law model is favored over the single power-law model by these factors.

We present our results in terms of the planetary mass ratio, $q$, instead of mass because
the mass ratio is measured robustly in virtually all planetary microlensing events. 
Converting mass ratios to masses requires the use of a Galactic model 
and an assumption about how the prevalence of planets might depend 
on Galactocentric distance and host star mass. The common assumption
is that there is no dependence on either, but this is motivated by our nearly complete ignorance of
these dependencies. By presenting our results in terms of the mass ratio, we avoid invalidating
them by making incorrect assumptions regarding the distance and host mass dependence of
the mass ratio function. Nevertheless, it is sometimes useful to consider what mass might
correspond to a particular mass ratio, such as the mass ratio of the break in the mass ratio
function.
Since our typical planet host star has a mass of $\sim 0.6\,M_\odot$, $q_{\rm br} \sim 10^{-4}$
corresponds to $\sim 20\,M_\oplus$, with an uncertainty of about a factor of two. 
This break in the mass ratio function seems to be at a somewhat larger mass than the breaks 
or peaks that appear
in {\it Kepler} populations, with planetary radii converted to masses following \citet{wol15}, but a larger
sample of low-mass microlens planets is needed to confirm this.

Our analysis is the most comprehensive systematic statistical analysis of the exoplanet 
occurrence frequency using planetary signals discovered in a microlensing survey. The
previous analyses relied on planet detections by follow-up groups, which have a lower
detection rate, and therefore, smaller samples. In the years since the MOA-II high cadence
survey started, several other high cadence surveys have commenced, including the OGLE-IV
survey \citep{ogle4}, the Wise survey \citep{shv15}, and the KMTNet survey \citep{kim16kmtnet}.

An even more powerful improvement to the statistical study of planetary systems found by
microlensing will be to include all the existing constraints on the mass and distance to the lens
systems. In particular, the lens system masses can often be measured when the microlensing
parallax effect is detected \citep{ben08,gau08,bennett-ogle109,mur11,sko14}, which can
be done much more easily with follow-up telescopes located far from the Earth  \citep{uda15,street16}.
For events without giant source stars, it is also possible to determine the lens system mass and
distance by detecting the lens star in high angular resolution follow-up observations
\citep{ben06,ben07,ben15hst,don09ob05071,bat14,bat15}. The inclusion of these constraints will
allow us, for the first time, to determine the dependence of the properties of cold exoplanets
as a function of host star mass and Galactocentric distance.

The ultimate word on the statistical properties of planetary systems will be 
achieved from the space based exoplanet survey
\citep{ben02} of the {\it WFIRST} \citep{spe15} mission, and hopefully also the {\it Euclid} \citep{pen13}
mission. The high angular resolution of these space telescopes will allow mass and distance
determinations of thousands of exoplanets because it will be possible to detect the lens star
and measure the lens-source relative proper motion with the high resolution survey data
itself. This will give us the same comprehensive picture of the properties of cold
exoplanets that {\it Kepler} is providing for hot planets.

\acknowledgments
D.S. and D.P.B. acknowledge support from NASA grants NNX13AF64G, NNX14AG49G,
and NNX15AJ76G.
T.S. acknowledges the financial support from the JSPS, JSPS23103002, JSPS24253004 and JSPS26247023. 
D.S. was supported by Grant-in-Aid for JSPS Fellows.
A.Y. acknowleges the financial support from the JSPS, JSPS25870893.
The MOA project is supported by the grant JSPS25103508 and 23340064.
This work was performed in part under contract with the California Institute of Technology (Caltech)/Jet Propulsion Laboratory (JPL) funded by NASA through the Sagan Fellowship Program executed by the NASA exoplanet Science Institute.
The authors thank OGLE, $\mu$FUN, MiNDSTEp, PLNAET, and RoboNet collaborations for letting us use their data to characterize
the alerted events.

\clearpage




\begin{table}
\caption{Selection Criteria for Well-monitored Events}
\label{tab:criteria}
 \begin{center}
  \begin{tabular}{ll}
   \tableline \tableline
   Cut-1 & $2 < t_{\rm E} < 300$ and $u_0 < 2$\\ \hline
   Cut-2 & $\sigma u_{0}/ u_{0} < 0.40$ or $\sigma u_{0} < 0.02$, $\sigma t_{\rm E} / t_{\rm E} < 0.25$ and $\sigma t_{\rm E} < 20$\\ \hline
   Cut-3 & $\sum {\rm f / fe} > 500$ within $\left| t-t_{0}\right| < t_{\rm E}$ \\ \hline
   Cut-4 & the number of data points, $N_{\rm data} > 30$ within $\left| t-t_{0}\right| < t_{\rm E}$ \\ \hline
   Cut-5 & $\sum {\rm f / fe} > 200$ within $0 < t-t_{0} < t_{\rm E}$  and $-t_{\rm E} < t-t_{0} < 0$\\ \hline
   Cut-6 & $N_{\rm data} > 10$ within $t - t_{0} < -20$ and $t - t_{0} > 20$\\ \hline
             & $N_{\rm data} > 3$ within $ |t - t_{0}| < u_{0} \times t_{\rm E}$ \\ 
   Cut-7 & $N_{\rm data} > 20$ within $|t - t_{0}| < 0.1 \times t_{\rm E}$ \\
             & $\sum {\rm f / fe} > 2000$ within $|t - t_{0}| < 0.2 \times t_{\rm E}$\\
   \tableline
  \end{tabular}
\vspace{-5mm}
\tablecomments{
The single lens model used for the event selection is described by three parameters, 
the time of the peak magnification, $t_{0}$, the Einstein radius crossing time(or event timescale), $t_{\rm E}$, 
and the impact parameter in units of the angular Einstein radius, $u_{0}$. The parameters with 
$\sigma$ in the table are 1-$\sigma$ uncertainties for the fit parameters. The parameter $\rm f$ and $\rm fe$ are the delta-flux (see text) and its error bar.
}
  \end{center}
\end{table}

\clearpage

\begin{table}
\caption{List of Planetary Events}
\label{tab:planet}
\begin{center}
{\small
\begin{tabular}{lccccccc}
\tableline \tableline
Event Name & $t_{\rm E}$ & $u_{0}$ & $q$ & $s$ & C.C. & OGLE ID & Reference \\
                       &    days &  $\theta_{\rm E}$ & $\times 10^{-3}$ & $\theta_{\rm E}$  \\
\tableline
MOA-2007-BLG-192 & 71.3 & 0.004 & 0.179 & 1.00 & ? & ---& (a) \\ 
MOA-2007-BLG-308 & 55.4 & 0.079 & 0.095 & 0.926 & yes & 368 & (b)  \\ 
MOA-2007-BLG-379 & 121 & 0.002 & 0.316 & 1.26 & yes & 349 & (c)  \\ 

MOA-2008-BLG-288 & 34.0 & 0.27 & 11.80 & 0.877 & yes  & 355 & (d) \\  
MOA-2008-BLG-379 & 42.5 & 0.0060 & 6.85 & 0.903 & yes & 570 & (e) \\  
                                  & 42.1 & 0.0060 & 6.99 & 1.119 & yes &        &  \\  

MOA-2009-BLG-266 & 61.45 & 0.13 & 0.058 & 0.91 & yes & --- & (f) \\ 
MOA-2009-BLG-319 & 16.6 & 0.0062 & 0.395 & 0.975 & yes   & --- & (g) \\  
MOA-2009-BLG-387 & 40.1 & 0.089 & 13.35 & 0.913 & yes & --- & (h) \\  

MOA-2010-BLG-117 & 116.3 & 0.279  &  0.75 & 0.86 & yes & --- & (i) \\ 
MOA-2010-BLG-328 & 62.6 & 0.072 & 0.26 & 1.15 & yes & --- & (j) \\ 
MOA-2010-BLG-353 & 11.1 & 0.5 & 1.38 & 1.46 & no & --- & (k) \\ 
MOA-2010-BLG-477 & 46.7 & 0.0034 & 2.18 & 1.12 & yes & --- & (l)  \\  

MOA-2011-BLG-028 & 34.4 & 0.908 & 0.181 & 1.673 & yes & 0203 & (m)  \\  
MOA-2011-BLG-197 & 53.63 & 0.130 & 3.954 & 1.04 & yes & 0265 & (n) \\  
MOA-2011-BLG-262 & 3.85 & 0.015 & 0.454 & 1.01 &  yes  & 0703 & (o) \\ 
MOA-2011-BLG-291 & 23.6 & 0.004 & 0.409 & 1.21 &  yes & --- & (p) \\ 
MOA-2011-BLG-322 & 23.17 & 0.046 & 28.4 & 1.822 &  no & 1127 & (q) \\  

MOA-2012-BLG-006 & 21.13 & 1.3 & 16.14 & 4.32  & yes & 0022 & (r) \\ 
MOA-2012-BLG-288 & 77.5 & 0.0014 & 1.09 & 0.41 & no & 0563 & (s) \\ 
                                  & 77.7 & 0.0014 & 1.09 & 2.43 & no &          &      \\ 
MOA-2012-BLG-355 & 22.2 & 0.19 & 0.70 & 0.76  &  yes & --- & (t) \\ 
MOA-2012-BLG-505 & 9.9 & 0.0076 & 0.19 & 0.91  &  yes & --- & (u) \\ 
                                  & 10.0 & 0.0075 & 0.20 & 1.12  &  yes &      &      \\ 
MOA-2012-BLG-527 & 67.0 & 0.103 & 0.24 & 0.890 &  no & 0950 & (v) \\ 
                                  & 66.5 & 0.104 & 0.22 & 1.004 &  no &          &       \\ 

\tableline
\end{tabular}
}
\vspace{-5mm}
\tablecomments{\scriptsize 
The C.C. column shows caustic crossing. \\
The OGLE ID column shows the last part of OGLE event name, i.e., OGLE-2007-BLG-368 for 368.\\
}
\vspace{-2mm}
\tablerefs{\scriptsize (a)\cite{ben08}; (b)\cite{sum10}; (c)\cite{ben16cir}; (d)\cite{kos14}; (e)\cite{suz14}; (f)\cite{mur11}; (g)\cite{miy11}; (h)\cite{bat11}; (i)\cite{ben16bin}; (j)\cite{fur13}; (k)\cite{rat15}; (l)\cite{bac12pl}; (m)\cite{sko15}; (n)\cite{sko14}; (o)\cite{ben14}; (p)\cite{ben17}; (q)\cite{shv14}; (r)Poleski et al. (in prep.); (s)\cite{fuk15}; (t)OGLE et al. (in prep.); (u)\cite{nag16}; (v)\cite{kos16};}
\end{center}
\end{table}
\clearpage

\begin{landscape}
\begin{table}
\caption{Parameters of the Ambiguous Event}
\label{tab:ambiguous}
\begin{center}
{\footnotesize
\begin{tabular}{lcccccccccccc}
\tableline \tableline
Event Name & Model & $t_{0} $ & $t_{\rm E}$ & $u_{0}$ & $q$ & $s$ & $\alpha$ & $t_*$ & $\chi^{2}$ & C.C. & OGLE ID\\ 
           &       & ${\rm HJD}^{\prime}$ & days & $10^{-3}$ & $10^{-3}$ &  & rad & days \\ 
\tableline
MOA-2011-BLG-336  & PL(close) 	& 5786.3980 	& -69.62 	& 8.05 & 0.611 & 0.688 & 4.741 & 0.296 & 8176.29 & no &  0950 \\ 
                  & $\sigma$ 	& 0.0006 	& 1.47  	& 0.17 & 0.038 & 0.011 & 0.003 & 0.006 & --- &     &	 \\ 
                  & PL(wide)  	& 5786.3974   	& 70.70 	& -7.95 & 0.551 & 1.409 & 1.600 & 0.301 & 8175.21 &  no   &	 \\ 
                  & $\sigma$ 	& 0.0007 	& 1.83  	& 0.21 & 0.035 & 0.021 & 0.003 & 0.005 & --- &     &	 \\ 
                  & SB(close) 	& 5786.3975  	& -65.53 	& -8.45 & 471.197 & 0.075 & 3.874 & 0.052 & 8196.07 &  no    &	 \\ 
                  & $\sigma$ 	& 0.0011 	& 1.21  	& 0.18 & 141.309 & 0.001 & 0.007 & 0.069 &---&     &	 \\ 
                  & SB(wide)  	& 5786.3940 	& -123.21 	& -4.49 & 2548.603 & 24.131 & 3.875 & 0.052 & 8195.01 &  no   &	 \\ 
                  & $\sigma$ 	& 0.0007 	& 7.70  	& 0.32 & 429.180 & 0.746 & 0.004 & 0.056 &---&     &	 \\
\tableline
\end{tabular}
}
\vspace{-5mm}
\tablecomments{\scriptsize
The C.C. column shows caustic crossing.\\
The first lines show the parameters of each model, and the second lines show the uncertainties. PL is planetary model, and SB is stellar binary model.\\
The OGLE ID column shows the last part of OGLE event name, i.e., OGLE-2011-BLG-0950.
}
\end{center}
\end{table}
\end{landscape}
\clearpage

\begin{table}
\caption{Best-fit Exoplanet Mass-ratio Function Parameters}
\label{tab:expl_bmrf}
\begin{center}
\begin{tabular}{lcccc}
\tableline \tableline
Parameter & \multicolumn{2}{c} {$q_{\rm br}$ Free} & \multicolumn{2}{c} {$q_{\rm br}$ Fixed} \\
                            &   MOA-only &  All    &   MOA-only &  All   \\
\tableline
$A$ &                                 $0.50$ & $0.62$ & $0.56$ & $0.61$ \\ 
$q_{\rm br}\times 10^4$ & $1.97$ & $1.65$ & $\equiv 1.7$ & $\equiv 1.7$ \\
$n$ &                                 $-0.96$ & $-0.92$ & $-0.94$ & $-0.92$ \\
$p$ &                                 $0.57$ & $0.47$ & $0.71$ & $0.44$ \\
$m$ &                                $0.65$ & $0.50$ & $0.65$ & $0.50$ \\
\tableline
\end{tabular}
\vspace{-2mm}
\end{center}
\end{table}

\begin{table}
\caption{Exoplanet Mass-ratio Function Parameters from MCMC}
\label{tab:expl_mrf}
\begin{center}
\begin{tabular}{lcccc}
\tableline \tableline
Parameter & \multicolumn{2}{c} {$q_{\rm br}$ Free} & \multicolumn{2}{c} {$q_{\rm br}$ Fixed} \\
                            &   MOA-only &  All    &   MOA-only &  All   \\
\tableline
$A$ & 				$0.65_{-0.26}^{+0.41}$ & $0.95_{-0.39}^{+0.48}$ & $0.56_{-0.17}^{+0.22}$ & $0.61_{-0.16}^{+0.21}$ \\ 
$B$ & 				$0.28^{+0.11}_{-0.08}$ & $0.29^{+0.09}_{-0.07}$ & $0.31^{+0.09}_{-0.08}$ & $0.34^{+0.09}_{-0.08}$ \\
$q_{\rm br}\times 10^4$ &	$1.19_{-0.58}^{+1.14}$ & $0.67_{-0.18}^{+0.90}$ & $\equiv 1.7$                    & $\equiv 1.7$ \\
$n$ & 				$-0.91_{-0.17}^{+0.15}$ & $-0.85_{-0.13}^{+0.12}$ & $-0.96_{-0.15}^{+0.14}$ & $-0.93 \pm 0.13$ \\
$p$ & 				$1.4_{-1.1}^{+3.4}$        & $2.6_{-2.1}^{+4.2}$        & $1.0_{-0.6}^{+0.9}$        & $0.6_{-0.4}^{+0.5}$ \\
$m$ & 				$0.62_{-0.58}^{+0.55}$ & $0.46 \pm 0.49$ & $0.62_{-0.58}^{+0.55}$    & $0.49_{-0.49}^{+0.47}$ \\
\tableline
\end{tabular}
\vspace{-5mm}
\tablecomments{Note that $B = f(q=3\times 10^{-4},s=1)$ is not an independent parameter,
but it has a smaller dispersion than $A$.
}
\vspace{-2mm}
\end{center}
\end{table}
\clearpage


\clearpage


\clearpage

\clearpage

\clearpage

\clearpage

\clearpage

\clearpage

\clearpage


\clearpage

\clearpage

\clearpage

\clearpage

\clearpage

\clearpage


\end{document}